\documentclass{emulateapj}

\begin{document}

\shortauthors{Luhman}

\shorttitle{Census of Stellar Populations in Sco-Cen}

\title{A Census of the Stellar Populations in the Sco-Cen Complex\altaffilmark{1}}

\author{K. L. Luhman\altaffilmark{2,3}}

\altaffiltext{1}
{Based on observations made with the Gaia mission, the Two Micron All
Sky Survey, the Wide-field Infrared Survey Explorer, and the eROSITA
instrument on the Spectrum-Roentgen-Gamma mission.}

\altaffiltext{2}{Department of Astronomy and Astrophysics,
The Pennsylvania State University, University Park, PA 16802, USA;
kll207@psu.edu}

\altaffiltext{3}{Center for Exoplanets and Habitable Worlds, The
Pennsylvania State University, University Park, PA 16802, USA}

\begin{abstract}

I have used high-precision photometry and astrometry from 
the early installment of the third data release of Gaia (EDR3) to
perform a survey for members of the stellar populations within the
Sco-Cen complex, which consist of Upper Sco, UCL/LCC, the V1062~Sco group,
Ophiuchus, and Lupus. Among Gaia sources with $\sigma_{\pi}<1$~mas,
I have identified 10,509 candidate members of those populations.
I have compiled previous measurements of spectral types, Li equivalent widths,
and radial velocities for the candidates, which are available for 3169, 1420,
and 1740 objects, respectively. 
In a subset of candidates selected to minimize field star contamination,
I estimate that the contamination is $\lesssim1$\% and the completeness is
$\sim90$\% at spectral types of $\lesssim$M6--M7 for the populations with
low extinction (Upper Sco, V1062~Sco, UCL/LCC).
I have used that cleaner sample to characterize the stellar populations in 
Sco-Cen in terms of their initial mass functions, ages, and space velocities. 
For instance, all of the populations in Sco-Cen have histograms of spectral 
types that peak near M4--M5, which indicates that they share similar 
characteristic masses for their initial mass functions 
($\sim0.15$--0.2~$M_\odot$).
After accounting for incompleteness, I estimate that the Sco-Cen complex
contains nearly 10,000 members with masses above $\sim0.01$~$M_\odot$.
Finally, I also present new estimates for the intrinsic colors of young
stars and brown dwarfs ($\lesssim20$~Myr) in bands from Gaia EDR3,
the Two Micron All Sky Survey, the Wide-field Infrared Survey Explorer, and 
the Spitzer Space Telescope.

\end{abstract}

\section{Introduction}
\label{sec:intro}

Scorpius-Centaurus \citep[Sco-Cen,][]{pm08} is the nearest OB association
to the Sun \citep[100--200~pc,][]{dez99}.
It contains several thousand stars with ages of 10--20~Myr
that historically have been divided into three subgroups: Upper Sco,
Upper Centaurus-Lupus (UCL), and Lower Centaurus-Crux 
\citep[LCC,][]{bla64,dez99}. More recent studies have suggested that 
UCL and LCC are two sections of a continuous distribution of stars
\citep{riz11} and have identified an additional 
compact group associated with V1062~Sco \citep{ros18,dam19}.
The subgroups of Sco-Cen overlap spatially and kinematically with hundreds
of younger stars ($\lesssim6$~Myr) associated with dark clouds
in Ophiuchus \citep{wil08} and Lupus \citep{com08}, which together comprise
the Sco-Cen complex\footnote{The Sco-Cen complex can be defined more
broadly to include additional clouds and young associations like Corona
Australis, Chamaeleon, and TW Hya that extend beyond the
Sco-Cen OB association \citep{pm08}.}.

The stellar populations in Sco-Cen are appealing for studies of star and
planet formation because they are nearby and contain large samples of stars
and circumstellar disks that span a range of ages and evolutionary stages.
However, the richest populations in Sco-Cen, Upper Sco and UCL/LCC,
are distributed across a large area of sky ($\sim60\arcdeg\times20\arcdeg$),
so a census of the complex requires wide-field data that can distinguish 
Sco-Cen members from numerous field stars. Discriminating among the 
overlapping populations in Sco-Cen poses an additional challenge.

The Gaia mission is performing an all-sky survey to measure precise 
photometry, proper motions, and parallaxes for more than a billion stars
\citep{per01,deb12,gaia16b}. Those data are extremely valuable for identifying
candidate members of nearby stellar associations. The first two data releases
of Gaia (DR1 and DR2) have been extensively utilized for that purpose in Sco-Cen
\citep{coo17,gol18,luh18,man18,ros18,wil18,can19,dam19,esp20,gal20b,mel20,luh20u,luh20lu,tei20,ker21}.
However, because of the complex kinematic structure of Sco-Cen
and varying approaches to membership analysis, some of the resulting samples
of candidates have differed substantially \citep[e.g.,][]{luh20lu}.

The early installment of the third data release of Gaia (EDR3) has improved
upon DR2 in terms of completeness and astrometric and photometric precisions
\citep{bro21}.
I have taken EDR3 as an opportunity to perform a thorough census of the
stellar populations in the Sco-Cen complex. I have used data from EDR3 to 
characterize the kinematics of the different populations within Sco-Cen 
and I have identified sources from EDR3 that have kinematics and photometry 
that support membership in those populations (Section~\ref{sec:ident}). 
I have compiled previous measurements of spectral types, Li equivalent
widths, and radial velocities for those candidates and I have examined the
stellar populations in Sco-Cen in terms of their initial mass functions (IMFs), 
ages, and space velocities (Section~\ref{sec:pop}). A separate study
identifies and classifies the circumstellar disks among the candidate members
of Sco-Cen \citep{luh21}.

\section{Identification of Candidate Members of Sco-Cen}
\label{sec:ident}

\subsection{Kinematic Selection Criteria}
\label{sec:kin}

For my survey of Sco-Cen, I have considered the area from $l=283$ to $2\arcdeg$
and $b=-12$ to $35\arcdeg$, which is large enough to extend beyond the outer
boundary of the Sco-Cen OB subgroups that was adopted by \citet{dez99}.

This study makes use of the following measurements from Gaia EDR3:
photometry in bands at 3300--10500~\AA\ ($G$), 3300--6800~\AA\ ($G_{\rm BP}$),
and 6300-10500~\AA\ ($G_{\rm RP}$); proper motions and parallaxes
($G\lesssim20$); radial velocities originating from Gaia DR2 ($G\sim4$--12);
and the renormalized unit weight error \citep[RUWE,][]{lin18}.
The latter serves as an indicator of the goodness of fit for the astrometry.
As done in my previous studies of Sco-Cen populations with Gaia DR2
\citep{esp20,luh20u,luh20lu}, I adopt a threshold of RUWE$<$1.6 when
I wish to consider astrometry from EDR3 that is likely to be reliable.

Ideally, the kinematic identification of members of an association would be
performed with three-dimensional space velocities. However, most stars in EDR3
lack measurements of radial velocities, which (in conjunction with proper
motions and parallactic distances) are needed for calculating space velocities.
Therefore, I rely on the proper motions and parallaxes from EDR3
for kinematic selection of candidates in Sco-Cen.
Because of projection effects, stars that share the same space velocity
but are distributed across a large area of sky can exhibit a broad range of
proper motions. To reduce such projection effects, I analyze the Gaia
astrometry in terms of a ``proper motion offset" ($\Delta\mu_{\alpha,\delta}$), 
which is defined as the difference between the observed proper motion of a
given star and the motion expected at the celestial coordinates and parallactic 
distance of the star for a characteristic space velocity for the association
(e.g., the median velocity), as done in my previous studies of Gaia data for
nearby associations \citep{esp17,esp19,esp20,luh18tau,luh20lu,luh20u}.
The proper motion offsets in this survey are calculated relative to the motions 
expected for a space velocity of $U, V, W = -5, -16, -7$~km~s$^{-1}$,
which approximates the median velocity of Upper Sco 
\citep[][Section~\ref{sec:uvw}]{luh20u}.
For parallactic distances, I adopt the geometric values estimated by 
\citet{bai21} from EDR3 parallaxes.

\begin{figure}
\epsscale{1}
\plotone{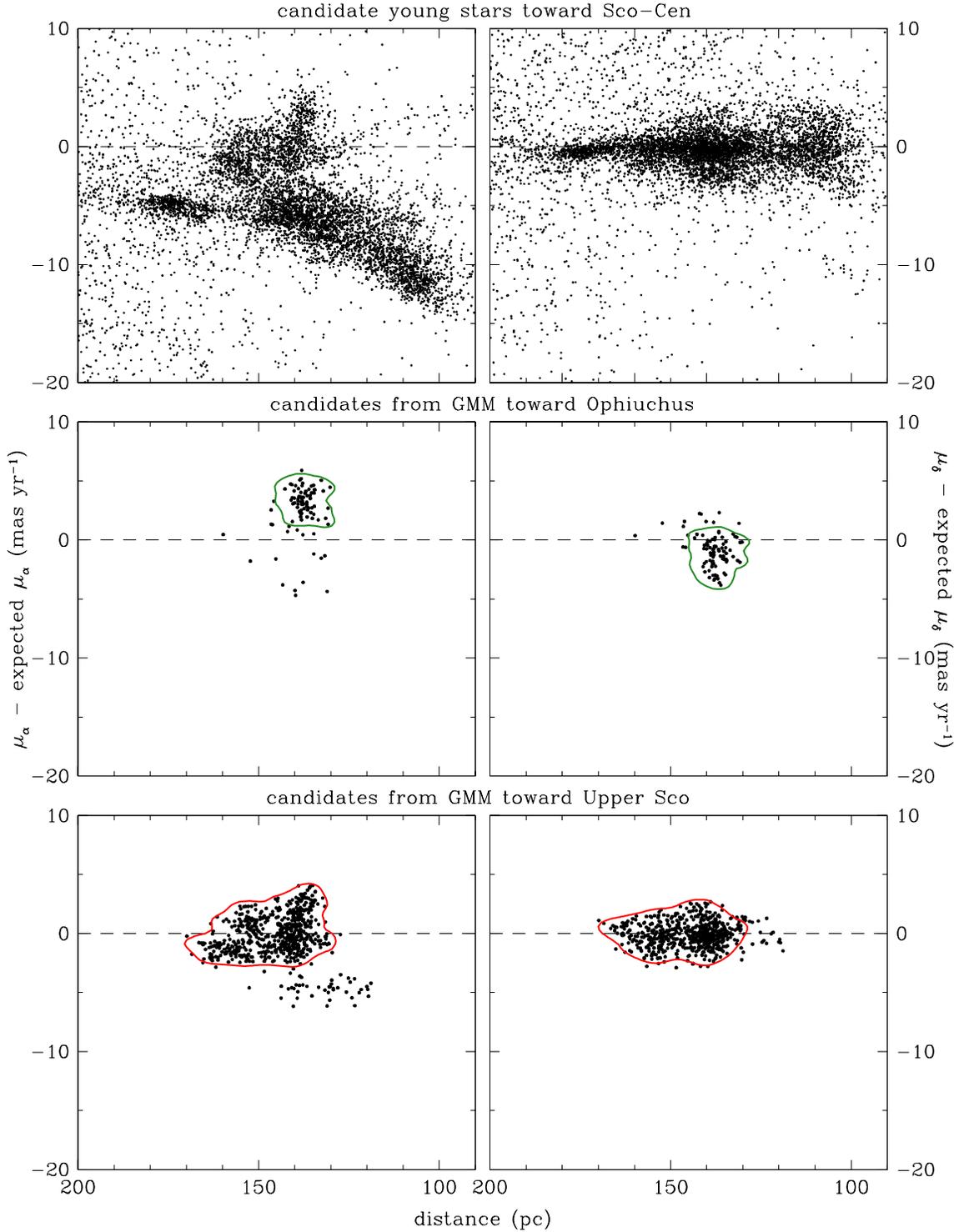}
\caption{
Proper motion offsets versus parallactic distance for candidate young low-mass
stars ($G_{\rm BP}-G_{\rm RP}$=1.4--3.4, $\sim$0.15--1~$M_\odot$) toward
Sco-Cen (top).
Candidate members of Sco-Cen have been identified by applying a GMM to these
data. I have plotted the resulting candidates that are located within
the boundary of Ophiuchus from \citet{esp18} (middle) and
the candidates that are outside of that boundary and within the triangular
area in Upper Sco from \citet{luh20u} (bottom).
I have included density contours that encompass the main concentration
of candidates in each sample, which serve as criteria for selecting candidate
members of these populations.
}
\label{fig:pp1}
\end{figure}

To characterize the kinematics of populations in Sco-Cen with Gaia EDR3,
I follow the procedures previously applied to data from Gaia DR2
for Upper Sco \citep{luh20u} and Lupus \citep{luh20lu}.
I begin by identifying candidates for young low-mass stars toward Sco-Cen by
selecting sources from Gaia EDR3 that have celestial coordinates within my
survey field, parallaxes within a range encompassing Sco-Cen ($\pi=4$--11~mas), 
small relative errors in parallax ($\pi/\sigma_{\pi}\geq20$), reliable
astrometry (RUWE$<$1.6), colors corresponding to low-mass stars 
($G_{\rm BP}-G_{\rm RP}$=1.4--3.4, $\sim$K5--M5, $\sim$0.15--1~$M_\odot$),
and positions above the single-star sequence for the Tuc-Hor association
\citep[45~Myr,][]{bel15} in $M_{G_{\rm RP}}$ versus $G_{\rm BP}-G_{\rm RP}$.
The resulting candidates are plotted in diagrams of proper motion offset 
versus parallactic distance in the top row of Figure~\ref{fig:pp1}.
Those data exhibit rich overlapping concentrations, which correspond
to the groups in Sco-Cen, and a sparse population that is roughly
uniform in density at a given distance, which consists of members
of the field (young stars and older unresolved binaries).
To discriminate between the Sco-Cen members and field stars, I have
applied a Gaussian mixture model (GMM) to the proper motion offsets and
distances using the {\tt mclust} package in R \citep{Rcore13,mclust}.
As done by \citet{luh20u} in analysis of DR2 data in Sco-Cen, I adopted
a model that contains three components for Sco-Cen (V1062~Sco, Upper 
Sco/Ophiuchus/Lupus, UCL/LCC) and a noise component for the field stars.
The stars with $>$90\% probability of membership in Sco-Cen according to
that model (i.e., the sum of the membership probabilities for the three
Sco-Cen components is $>$90\%) are adopted as an initial sample of candidate 
members of Sco-Cen for the following analysis.

The GMM has provided a means of separating candidate members of Sco-Cen
from field stars. To isolate the kinematics of the individual populations in
Sco-Cen, I can examine the candidates within the areas where
the populations are concentrated, which are revealed by
my Sco-Cen candidates as well as previous surveys of Sco-Cen.
The Upper Sco association is projected against the Ophiuchus clouds and extends
well beyond them \citep{dez99}.
\citet{esp18} attempted to identify the area within which Ophiuchus members
dominate by using Gaia DR2 data to measure the spatial variation of
relative ages of stars in the vicinity of Ophiuchus.
They defined a boundary that encompassed stars with systematically younger
ages, which roughly coincided with the edges of the Ophiuchus clouds.
Meanwhile, \citet{luh20u} defined a triangular field that contains 
the central concentration of stars in Upper Sco (and Ophiuchus).
In \citet{luh20lu}, I found that most of the members of Lupus are
within four fields that encompass clouds 1--4 \citep[see also][]{gal20b}.
The members of the V1062~Sco group are clustered within a few
degrees of that star \citep{ros18,dam19,luh20u}. 
Finally, based on the candidates identified with my GMM and similar work
with DR2 data in \citet{luh20u}, members of UCL/LCC exhibit the widest
spatial distribution among the populations in Sco-Cen, extending across
the entire length of the OB subgroups as defined by \citet{dez99}.

\begin{figure}
\epsscale{1}
\plotone{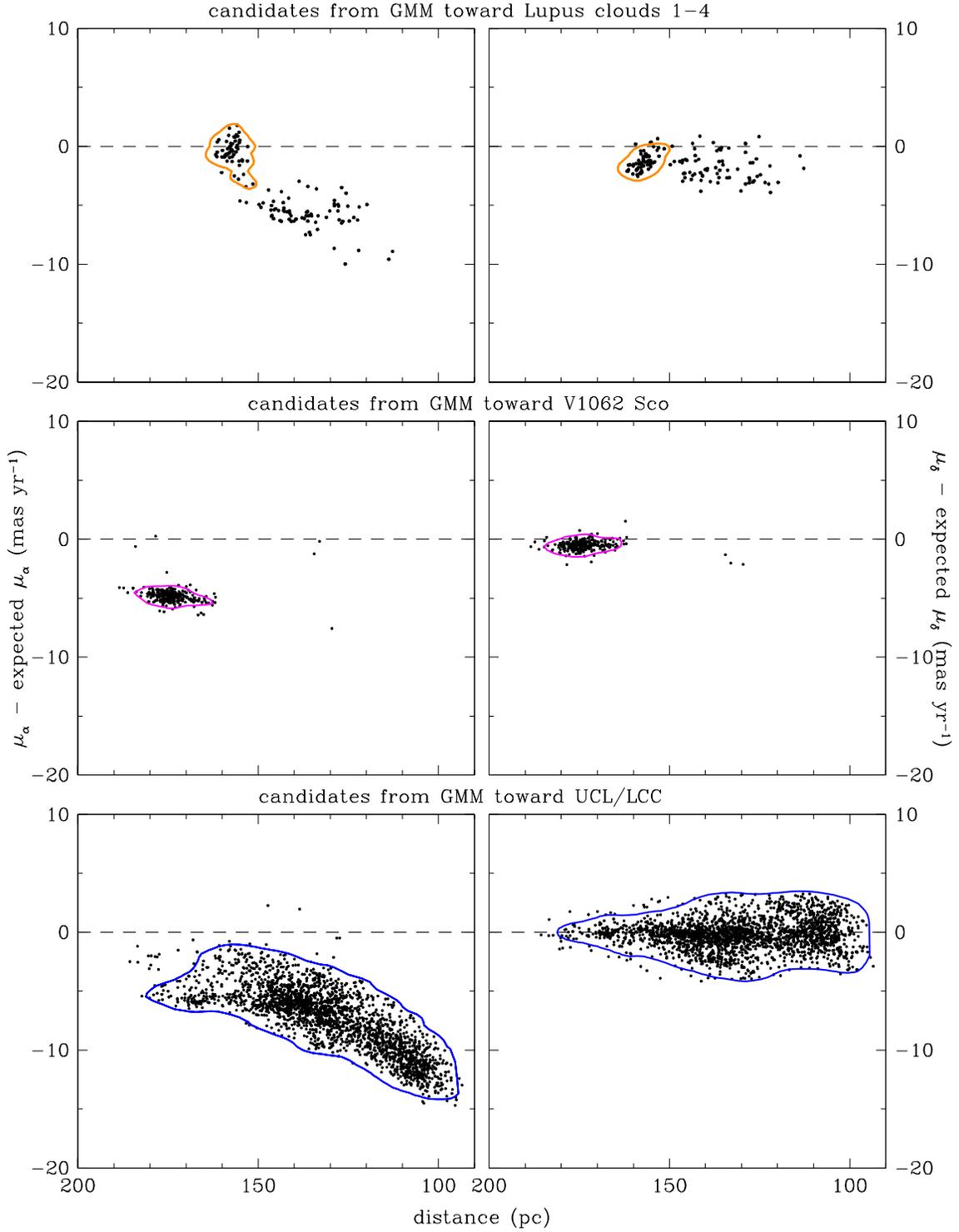}
\caption{
Same as Figure~\ref{fig:pp1} but for Lupus clouds 1--4 (top), 
the V1062~Sco group (middle), and UCL/LCC (bottom).
}
\label{fig:pp2}
\end{figure}

In each of five panels in Figures~\ref{fig:pp1} and 
\ref{fig:pp2}, I have plotted the proper motion offsets and parallactic 
distances of all Sco-Cen candidates identified with the 
GMM that are located within the following five regions:
1) the Ophiuchus field from \citet{esp18};
2) the triangular field in the center of Upper Sco from \citet{luh20u},
excluding Ophiuchus;
3) fields toward Lupus clouds 1--4 from \citet{luh20lu}; 
4) a $2\arcdeg$ radius field toward V1062~Sco;
and 5) a field encompassing most of the stars in the UCL/LCC component from
the GMM, which I define as $l=295$--$312\arcdeg$/$b=0$--$20\arcdeg$ and
$l=312$--$343\arcdeg$/$b=10$--$20\arcdeg$.
In these data, the kinematics of the individual populations are readily
identified. Each of the diagrams for V1062~Sco and UCL/LCC exhibits a single
well-defined cluster. The former is compact and the latter is much broader,
particularly in distance, which is consistent with the relative sizes
of these populations on the sky.
I have used the functions {\tt kde2d} and 
{\tt bandwidth.nrd} from the {\tt MASS} package in R \citep{mass}
to calculate density maps in $(\pi,\Delta\mu_{\alpha})$ and
$(\pi,\Delta\mu_{\delta})$ for V1062~Sco and UCL/LCC and I have marked
the resulting contours that encompass most ($>95$\%) of the candidates
in each cluster. For each of the fields toward Ophiuchus and Upper Sco,
the kinematic data show one dominant cluster that corresponds to
the targeted population and a smaller number of outliers. 
For the Ophiuchus field, the data for the outliers are consistent with 
membership in Upper Sco or UCL/LCC. The outliers in the Upper Sco field
are likely members of UCL/LCC.
In the kinematic diagram for the Lupus fields, the most compact group
contains stars associated with the clouds while the remaining stars are
likely members of UCL/LCC \citep{luh20lu}. As done for V1062~Sco and UCL/LCC,
I have marked density contours in Figures~\ref{fig:pp1} and \ref{fig:pp2}
for the clusters that correspond to members of Ophiuchus, Upper Sco, and Lupus.

I have used the contours in Figures~\ref{fig:pp1} and \ref{fig:pp2}
as criteria for identifying candidate members of the Sco-Cen populations.
I have retrieved from EDR3 all sources that have positions within my survey
field, parallax errors of $\sigma_{\pi}<1$~mas, and distances and proper
motion offsets that overlap at 1~$\sigma$ with any of the five pairs of
contours. A criterion involving RUWE is not applied.
This sample will be further refined with Gaia photometry in the
next section.

I note that the selection of candidates using the contours in
Figures~\ref{fig:pp1} and \ref{fig:pp2} has negligible dependence on the 
precise value of the velocity assumed when calculating the proper motion 
offsets. The populations in Sco-Cen span a fairly small range of space 
velocities (5--10~km~s$^{-1}$, Section~\ref{sec:uvw}), so adopting a different
velocity within that range (e.g., the median velocity of UCL/LCC instead
of the median velocity of Upper Sco) would result in a small shift in the
proper motion offsets (and their contours) that is nearly the same for all
Sco-Cen members, leaving the selection of candidates unaffected.

\subsection{Photometric Selection Criteria}
\label{sec:phot}

\begin{figure}
\epsscale{1.1}
\plotone{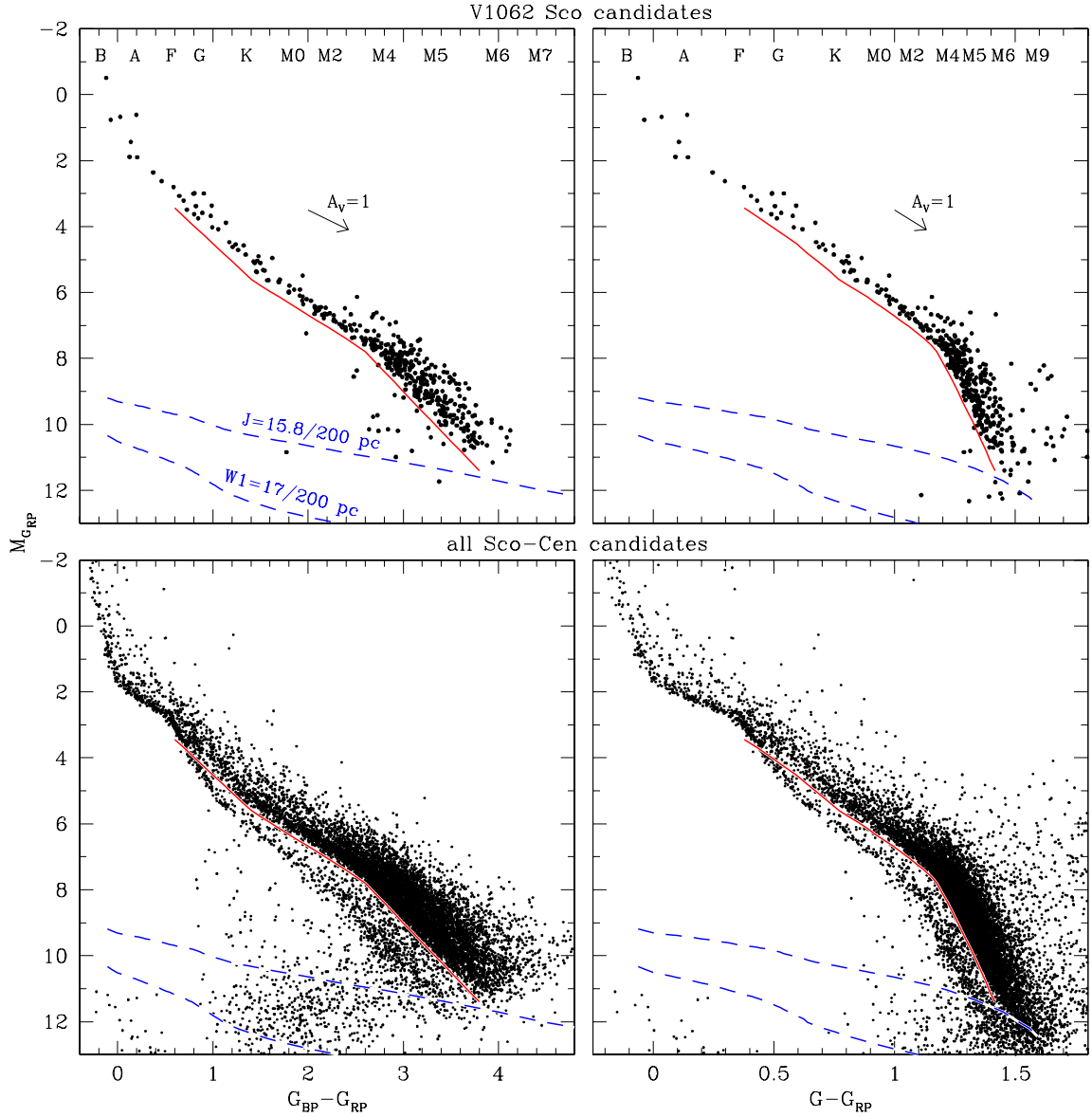}
\caption{
$M_{G_{\rm RP}}$ versus $G_{\rm BP}-G_{\rm RP}$ and $G-G_{\rm RP}$ 
for candidate members of V1062~Sco within a radius of $2\arcdeg$ from
the center of that group (top) and all candidate members of Sco-Cen (bottom) 
selected with the kinematic criteria in Figures~\ref{fig:pp1} and \ref{fig:pp2}.
To further refine those kinematic candidates, I have defined boundaries that
follow the lower envelopes of the sequences in V1062~Sco (red solid lines)
and I have rejected candidates that appear below them unless they exhibit 
mid-IR excess emission, in which case they are retained.
For reference, I have indicated the spectral types that 
correspond to the colors of young stars (Table~\ref{tab:intrinsic})
and I have marked the completeness limits for 2MASS and WISE 
in $J$ and W1, respectively, based on those colors and a distance of 200~pc,
which is just beyond the far side of Sco-Cen (blue dashed lines).
}
\label{fig:cmd}
\end{figure}

The candidate members of Sco-Cen selected via kinematics in the previous
section can be further refined with color-magnitude diagrams (CMDs)
consisting of $M_{G_{\rm RP}}$ versus $G_{\rm BP}-G_{\rm RP}$
and $M_{G_{\rm RP}}$ versus $G-G_{\rm RP}$. 
Since the V1062~Sco group and UCL/LCC are tied as the oldest populations
in Sco-Cen \citep{luh20u}, any members of Sco-Cen should appear on or above
the sequences of those populations in CMDs.
The sequence for V1062~Sco is more easily measured since that group is much
more compact than UCL/LCC.
Selecting candidates within a small area that encompasses
the bulk of the V1062~Sco group produces a sample that has little
contamination from field stars or other populations in Sco-Cen.
In the top row of Figure~\ref{fig:cmd}, I have plotted two CMDs for the
candidates with kinematics consistent with V1062~Sco and positions within
a radius of $2\arcdeg$ from the center of that group.
Photometry with errors greater than 0.1~mag has been excluded.
As expected, the sequences for V1062~Sco are clearly defined.
I have marked boundaries in Figure~\ref{fig:cmd} that follow the lower
envelopes of those sequences.

In each of the two CMDs in the bottom row of Figure~\ref{fig:cmd}, I have
plotted all kinematic candidates in Sco-Cen from the previous section that have
photometry with errors less than 0.1~mag in both bands for a given CMD.
I have included the boundaries defined with the sequences for V1062~Sco,
and I have rejected candidates that appear below either of them
unless they exhibit infrared (IR) excess emission in photometry from
the Wide-field Infrared Survey Explorer \citep[WISE,][]{wri10}.
Stars with excesses have been identified using colors within
the WISE bands \citep[W1$-$W2, W1$-$W3, W1$-$W4,][]{luh21}.
The latter candidates are retained since stars that are occulted by
circumstellar disks can appear in scattered light, which results in 
unusually faint apparent magnitudes at a given color.
When rejecting candidates with the CMD boundaries, the absolute magnitudes
of the candidates are calculated for the 1~$\sigma$ upper limit on the distance
from \citet{bai21} (i.e., the brightest $M_{G_{\rm RP}}$ allowed at 1~$\sigma$).
If that upper limit is greater than 200~pc, a value of 200~pc is adopted since
it approximates the maximum distance for members of Sco-Cen 
(Figure~\ref{fig:pp1}).
The number of candidates satisfying at least one CMD and not rejected by either
CMD is 10,104. The number of additional candidates
that fell below the boundary of a CMD but were
retained as viable candidates because of IR excess is 91.

Some kinematic candidates were not plotted in either of the CMDs
in Figure~\ref{fig:cmd} because photometry is unavailable 
or uncertain in one or more Gaia bands. In most cases, this is due to
the faintness of a star or its close proximity to a brighter star.
The former tend to have large astrometric uncertainties (due to their faint
magnitudes) while the latter tend to be brighter and have more precise
astrometry. A natural division between these two groups appears
near $\sigma_{\pi}\sim0.4$~mas. Among these candidates that lack
photometry for CMDs, I have rejected those with $\sigma_{\pi}>0.4$~mas
except for two objects that have IR excess emission, which are retained.
I also have rejected candidates with $\sigma_{\pi}<0.4$~mas if they are
within $5\arcsec$ from a star that shares a similar proper motion and parallax
and that is rejected by the CMDs (i.e., a likely companion to a rejected
star). The remaining 312 candidates that lack the data for CMDs and
have $\sigma_{\pi}<0.4$~mas are retained.

The 10,509 candidates are presented in Table~\ref{tab:cand}. The spatial 
distribution of the kinematic populations among these candidates is 
illustrated in Figure~\ref{fig:map}.
For reference, I have marked in those maps the boundary
of Sco-Cen from \citet{dez99} and the fields in Upper Sco, Ophiuchus,
Lupus, and V1062~Sco that were mentioned in Section~\ref{sec:kin}.
Since the kinematics of the populations in Sco-Cen overlap 
(Figure~\ref{fig:pp1}), data for a given candidate can be consistent with
multiple populations. Two of the maps in Figure~\ref{fig:map} show the
candidates with kinematics consistent with Upper Sco, Ophiuchus, or Lupus (top)
and the candidates consistent with V1062~Sco (middle). To better trace the
spatial distribution for UCL/LCC and minimize contamination from the other
populations, the third map shows candidates with kinematics indicative of 
only UCL/LCC and no other population.
The latter map illustrates how UCL/LCC extends across Lupus, Ophiuchus,
and Upper Sco. As a result, pre-Gaia surveys for members of the latter
regions have been subject to contamination from UCL/LCC 
\citep{gal20b,luh20u,luh20lu}.

\subsection{Non-kinematic Candidate Companions}

I have attempted to identify companions to the candidates in
Table~\ref{tab:cand} that are in Gaia EDR3 but did not satisfy
the kinematic criteria that produced those candidates.
To do that, I retrieved sources from EDR3 that are located within $5\arcsec$ 
from the candidates and that satisfy any of these three sets of criteria:
1) a parallax measurement is unavailable or uncertain ($\sigma_{\pi}>1$~mas)
and the boundaries in both CMDs in Figure~\ref{fig:cmd} are satisfied
for a distance of 200~pc (the most optimistic distance to adopt);
2) the parallax error is below 1~mas, the kinematics do not satisfy any of the 
criteria from Figures~\ref{fig:pp1} and \ref{fig:pp2}, and both CMDs are 
satisfied for 200~pc
or IR excess emission is present (all of these sources share similar
proper motions with their potential companions from Table~\ref{tab:cand});
3) photometry for CMDs is unavailable, the kinematics do not satisfy any of the 
criteria from Figures~\ref{fig:pp1} and \ref{fig:pp2}, and the parallax and 
proper motion are within roughly similar to those of the candidate from 
Table~\ref{tab:cand} that it is near ($\Delta\pi<2$~mas, 
$\Delta\mu_{\alpha,\delta}<5$~mas).
The resulting samples contain 48, 100, and 47 sources, respectively.
These 195 candidate companions are presented in Table~\ref{tab:comp}.
Many of the candidates have large values of RUWE that would suggest unreliable
astrometry, which could explain why they did not satisfy the kinematic criteria
for membership.

\subsection{Compilation of Data for Candidates}

In Tables~\ref{tab:cand} and \ref{tab:comp}, I have compiled various
data that are useful for analyzing the candidate members of Sco-Cen.
All candidates were selected from Gaia EDR3, so all have source names
from that catalog. To facilitate comparison to previous work, I have included
the names for matching sources from Gaia DR2 that are within a separation 
of $0\farcs1$ and the designations from a selection of other catalogs.
The remaining contents of Tables~\ref{tab:cand} and \ref{tab:comp}
consist of the equatorial coordinates, proper motion, parallax, RUWE,
and photometric magnitudes from Gaia EDR3; measurements of
spectral types and the types adopted in this work; the average of previous 
measurements of equivalent widths for Li at 6707~\AA; 
distance estimate based on Gaia EDR3 parallax \citep{bai21}; the most 
accurate radial velocity measurement from 
previous studies that has an error less than 4~km~s$^{-1}$; 
$UVW$ velocities (Section~\ref{sec:uvw}; only for
Table~\ref{tab:cand}); a flag indicating the populations with which
the kinematics are consistent based on the criteria in
Figures~\ref{fig:pp1} and \ref{fig:pp2} (only for Table~\ref{tab:cand});
a flag indicating the regions of Sco-Cen in which a candidate is located;
the designations and angular separations of the closest sources within
$3\arcsec$ from the Point Source Catalog of the Two Micron All Sky Survey
\citep[2MASS,][]{skr06} and the WISE All-Sky Source Catalog \citep{cut12},
the AllWISE Source Catalog, or the AllWISE Reject Catalog \citep{cut13a};
and flags indicating whether the Gaia source is the closest match
in EDR3 for the 2MASS and WISE sources.
The numbers of candidates in Table~\ref{tab:cand} that have
measurements of spectral types, Li, and radial velocities are
3169, 1420, and 1740, respectively. Three additional objects have
been observed spectroscopically, but their classifications are uncertain
(i.e., no adopted spectral type in Table~\ref{tab:cand}).
For close pairs that were likely unresolved during previous spectroscopy,
the spectroscopic data have been assigned to the component that is brighter
in $G$ from EDR3. For the closest matching WISE sources, \citet{luh21} provides 
a catalog of IR photometry from 2MASS and WISE, flags indicating whether those
data exhibit excess emission from disks, and classifications of the
evolutionary stages of any detected disks.

\subsection{Comparison of Gaia, 2MASS, and WISE Limits}

Given that 2MASS and WISE are the primary sources of IR photometry for
the candidate members of Sco-Cen, it is useful to compare their limits to
that of Gaia EDR3. For this comparison, I have chosen the bands from 2MASS
and WISE that are most sensitive to late-type members of Sco-Cen, which 
correspond to $J$ and W1 given the colors of such objects and the relative 
completeness limits among the 2MASS and WISE bands 
\citep{skr06,cut13b,eis20,mar21}.
The completeness limits in $J$ and W1 are $\gtrsim15.8$ and 17, respectively,
for most of the sky. In the CMDs in Figure~\ref{fig:cmd}, I have marked
the values of $M_{G_{\rm RP}}$ that correspond to those limits assuming the
typical colors of young stars (Section~\ref{sec:spt}) and a distance just
beyond the far side of Sco-Cen (200~pc). Assuming smaller distances would move
the limits down in those diagrams. A comparison of the $J$ and W1 limits and 
the sequences for Sco-Cen indicates the following for members of Sco-Cen:
WISE is deeper than 2MASS; 2MASS and WISE are deeper than $G_{\rm BP}$;
2MASS probably has a roughly similar depth as $G$ and $G_{\rm RP}$.
All Sco-Cen members with $\sigma_{\pi}<1$~mas (the threshold used for
the candidates in Figure~\ref{fig:cmd}) should be detected in W1 as
long as they are not blended with another star.

\subsection{Field Star Contamination}
\label{sec:field}

Additional criteria can be applied to the candidate members of Sco-Cen
identified in Sections~\ref{sec:kin} and \ref{sec:phot} (Table~\ref{tab:cand})
to minimize contamination from field stars.
The map of the candidates in Figure~\ref{fig:map} demonstrates that all
of the populations in Sco-Cen are concentrated within the boundaries of the
OB subgroups from \citet{dez99}. Therefore, considering only candidates within
that area reduces the relative contribution of field stars to the sample.
Contamination is further reduced by requiring reliable astrometry
(RUWE$<$1.6) and $\sigma_{BP}<0.1/\sigma_{RP}<0.1$ or
$\sigma_G<0.1/\sigma_{RP}<0.1$ 
so that the candidates have satisfied at least one of the two CMDs 
in Figure~\ref{fig:cmd}.

I have estimated the field star contamination in the subset of candidates
from Table~\ref{tab:cand} that satisfy the preceding criteria. 
To do that, I shifted the kinematic thresholds from Figures~\ref{fig:pp1}
and \ref{fig:pp2} by +20~mas~yr$^{-1}$ in $\Delta\mu_{\alpha}$ and 
+12~mas~yr$^{-1}$ in $\Delta\mu_{\delta}$, retrieved sources from EDR3 
within the Sco-Cen survey field that satisfied those shifted thresholds, and 
applied the CMD boundaries in Figure~\ref{fig:cmd} to the resulting kinematic 
candidates. Those steps were repeated for seven additional shifts that surround
the thresholds for Sco-Cen: (+20,0), (+20,$-$12), (0,+12), (0,$-$12), 
($-$20,+12), ($-$20,0) and ($-$20,$-$12)~mas~yr$^{-1}$. 
The sizes of the shifts were selected to be large enough that
the new thresholds did not overlap with the kinematics of Sco-Cen.
In the eight resulting samples of field stars
that satisfied the CMD criteria for Sco-Cen, the numbers of stars range from
$\sim40$--100 for $\sigma_{\pi}<0.5$~mas, which is a small variation
relative to the thousands of candidates in Sco-Cen. However, the variation
in the numbers of stars is greater for $\sigma_{\pi}>0.5$~mas and the highest
numbers occurred for the kinematic thresholds that were closest to the 
proper motion offsets exhibited by the bulk of distant field stars beyond
Sco-Cen. A small fraction of those numerous distant stars can have parallax
measurements within the range for Sco-Cen due to their large errors,
enabling their selection as kinematic candidates for Sco-Cen. When their
absolute magnitudes are calculated using the erroneously small parallactic
distances, they appear anomalously faint in CMDs. Stars of this kind
comprise the large clump below the main sequence in the 
$G_{\rm BP}/G_{\rm RP}$ CMD for Sco-Cen candidates in Figure~\ref{fig:cmd}.
In the $G/G_{\rm RP}$ CMD, the clump straddles the bottom of the main sequence.
When those stars have photometry in $G_{\rm BP}$ and $G_{\rm RP}$, they
are rejected by that CMD.
But for the clump stars that only have $G$ and $G_{\rm RP}$, many of them
appear above the boundary in that CMD and survive to contaminate the sample
of candidates. This source of contamination can be greatly reduced by
considering only candidates with $\sigma_{\pi}<0.5$~mas, 
which corresponds to 10,145 of the 10,509 candidates in Table~\ref{tab:cand}.

\begin{figure}
\epsscale{1.2}
\plotone{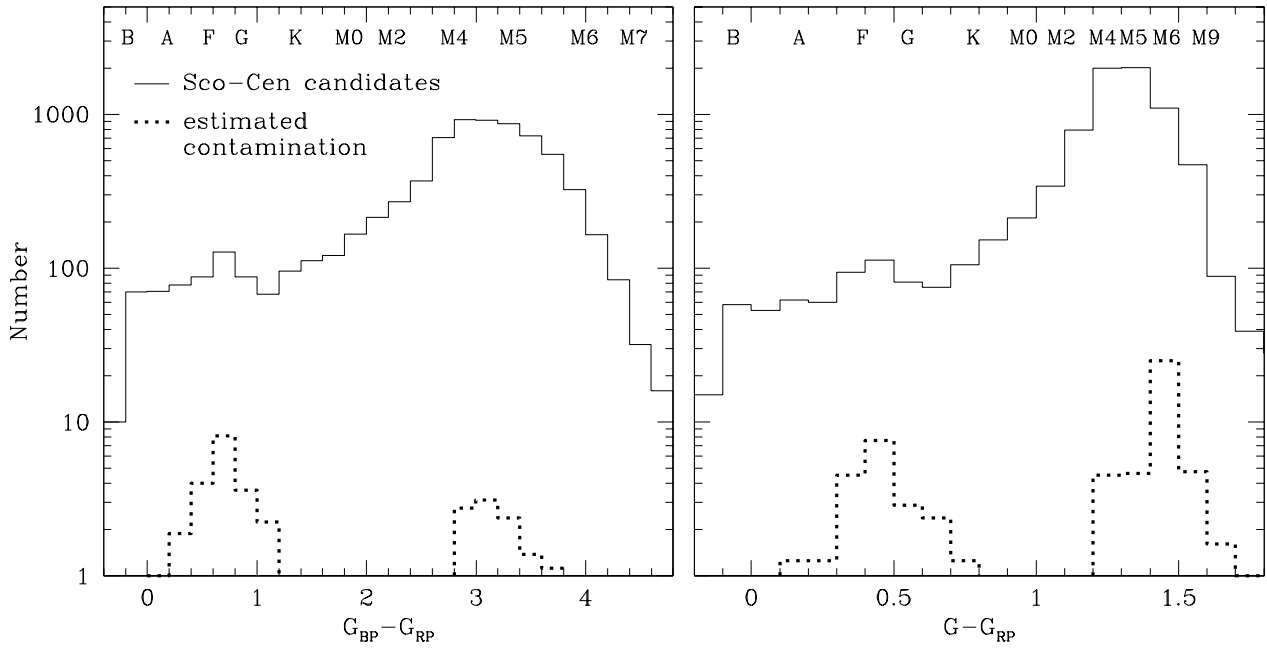}
\caption{
Histograms of $G_{\rm BP}-G_{\rm RP}$ and $G-G_{\rm RP}$ for the most reliable
sample of Sco-Cen candidates (solid lines)
and the estimated contamination from field stars (dotted lines).
For each color, the candidates are required to have photometric errors
of $<0.1$~mag in both bands, RUWE$<$1.6, $\sigma_{\pi}<0.5$~mas, and locations
within the Sco-Cen boundaries from \citet{dez99}.
}
\label{fig:off}
\end{figure}

In Figure~\ref{fig:off}, I have plotted histograms of $G_{\rm BP}-G_{\rm RP}$
and $G-G_{\rm RP}$ for Sco-Cen candidates from Table~\ref{tab:cand}
that satisfy the criteria described thus far in this section for minimizing
contamination: locations within the Sco-Cen boundary from \citet{dez99},
RUWE$<$1.6, $\sigma_{\pi}<0.5$~mas, and $\sigma_{BP}<0.1/\sigma_{RP}<0.1$ or
$\sigma_G<0.1/\sigma_{RP}<0.1$.
The resulting numbers of candidates in the two histograms are 7272 and 8000,
respectively. For comparison, I have included in Figure~\ref{fig:off} the 
average histograms for the eight samples of field stars described 
previously after applying the same criteria for minimizing field stars. 
The resulting contaminants consist of A/F/G stars on the main sequence
and M stars appearing above the main sequence. The former source of
contamination is expected given that the CMD boundaries for selecting candidates
intersect with the main sequence (Figure~\ref{fig:cmd}). The group of M-type
field likely consists of both young stars and older unresolved binaries.
The field star histograms for $G_{\rm BP}/G_{\rm RP}$ and $G/G_{\rm RP}$
contain $\sim$40 and 70 stars, respectively, which correspond to $\sim$0.6\%
and 0.9\% of the Sco-Cen candidates. 
Requiring accurate photometry in $G_{\rm BP}$ and $G_{\rm RP}$ produces
the cleanest sample of candidates, but the sample that only requires 
$G$ and $G_{\rm RP}$ reaches later spectral types.

\begin{figure}
\epsscale{1}
\plotone{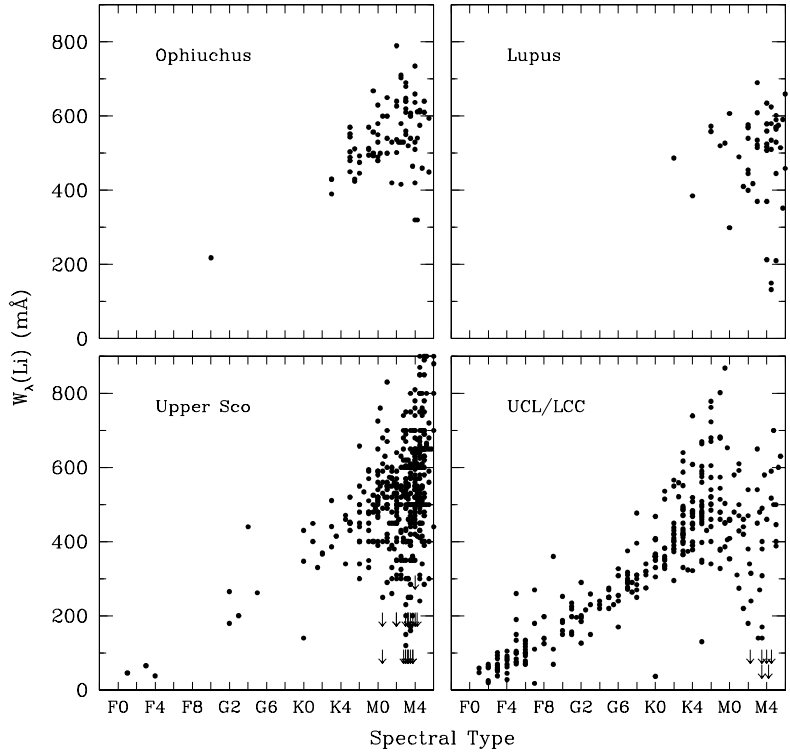}
\caption{
Equivalent widths of Li versus spectral type for candidate members of
Ophiuchus, Lupus, Upper Sco, and UCL/LCC.
}
\label{fig:li}
\end{figure}

Spectroscopy can be used to assess whether individual candidates are likely
to be field stars or members of Sco-Cen.
Measurements of radial velocities can further constrain membership by
enabling the calculation of $UVW$ velocities, although multiple epochs
are necessary to check whether the velocities are affected by orbital
motion within binary systems.
The Li absorption line at 6707~\AA\ and gravity-sensitive spectral features
serve as diagnostics of youth that can distinguish between members of Sco-Cen
and older field stars.
To illustrate the application of Li for this purpose, I have plotted in
Figure~\ref{fig:li} the Li equivalent widths compiled in Table~\ref{tab:cand}
versus spectral type for the samples in Ophiuchus, Lupus, Upper Sco,
and UCL/LCC defined in Section~\ref{sec:clean}. The sample for V1062~Sco
has been omitted since few of its candidates have Li data.
In low-mass stars, Li is depleted over time at a
rate that varies with spectral type \citep{bil97}.
In Figure~\ref{fig:li}, the candidates that exhibit unusually weak Li relative 
to other candidates at a given spectral type may be older field stars.
It would be useful to better test the membership of those stars with
weak Li by searching for evidence of youth with gravity-sensitive features and
comparing their $UVW$ velocities to those of other members. In addition,
measuring Li for a larger number of M-type candidates in UCL/LCC and 
the V1062~Sco group would provide constraints on their ages 
\citep{sta99,men08,yee10}, which could be compared to the ages inferred from 
their CMDs (Section~\ref{sec:ages}).

\subsection{Completeness}
\label{sec:completeness}

In the previous section, I found that field star contamination among the
candidate members of Sco-Cen is reduced by considering only candidates that
have RUWE$<$1.6, $\sigma_{\pi}<0.5$~mas, and $\sigma_{BP}<0.1/\sigma_{RP}<0.1$ 
or $\sigma_G<0.1/\sigma_{RP}<0.1$.
These criteria will be adopted when characterizing
the stellar populations in Sco-Cen in Section~\ref{sec:pop}.
Therefore, it is useful to estimate the completeness of the samples of
candidates defined in that way.

\begin{figure}
\epsscale{0.5}
\plotone{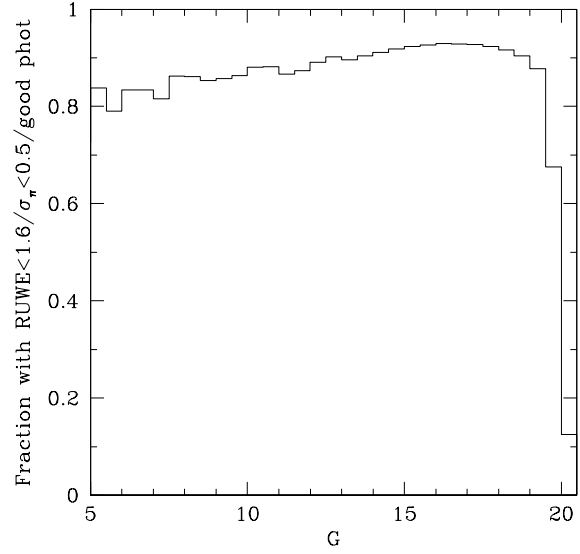}
\caption{
Fraction of sources from Gaia EDR3 toward Sco-Cen that have RUWE$<$1.6,
$\sigma_{\pi}<0.5$~mas, and $\sigma_{BP}<0.1/\sigma_{RP}<0.1$ or
$\sigma_G<0.1/\sigma_{RP}<0.1$.
The data in this diagram provide an estimate of the completeness of samples
of candidates in Sco-Cen that are required to satisfy those criteria in
order to reduce field star contamination (Sections~\ref{sec:field} and
\ref{sec:clean}).
}
\label{fig:completeness}
\end{figure}

Gaia EDR3 has a high level of completeness at $G\lesssim19$--20 except for
very crowded regions and the brightest stars that experience saturation 
\citep{fab21}. $G$-band measurements are available for $>99$\% of the
sources in EDR3 \citep{bro21}.
For sources with data in $G$ that are within the boundary of Sco-Cen
from \citet{dez99}, I have plotted the fraction that satisfy the
previously listed criteria for reducing field star contamination
versus $G$ in Figure~\ref{fig:completeness}. 
The fraction is $\sim85$\% at $G<12$ and $\sim90$--95\% at $G=12$--19.5.
The lower values at brighter magnitudes are due to the restriction on RUWE.
These data indicate that the cleanest samples of candidate members of
Sco-Cen may be missing $\sim10$\% of the members. Note that some of those
missing members are likely among the candidates in Table~\ref{tab:cand}
(and Table~\ref{tab:comp}), but they are in regimes of RUWE, $\sigma_{\pi}$,
and photometric errors in which field star contamination is higher.

\subsection{Comparison to Previous Surveys}
\label{sec:compare}

Table~\ref{tab:prev} lists many of the studies from the last two decades
that have identified candidate members of populations 
in Sco-Cen or have studied samples of candidates from other sources. 
Most surveys of Lupus have been omitted since they were scrutinized with
Gaia DR2 by \citet{luh20lu} and the results have not changed qualitatively
with EDR3. For each study in Table~\ref{tab:prev}, I have compiled the number 
of candidate members in its sample, the number of those candidates that
have parallax measurements with $\sigma_{\pi}<1$~mas from EDR3, and the
number of the candidates with parallaxes that have been classified in this work
as candidate members of any Sco-Cen population (i.e., they appear in
Table~\ref{tab:cand}).

Because of saturation, Gaia EDR3 is increasingly incomplete at decreasing
magnitudes below $G\sim4$ \citep{gaia16b,fab21}.
As a result, many of the brightest stars that have been previously
proposed as members of Sco-Cen (e.g., Antares) are absent from my catalog
of candidates.
For instance, 14 of the candidates from \citet{dez99} lack parallax
measurements from EDR3 and have Hipparcos magnitudes of $\lesssim4$
\citep{van07}. The analysis of the main sequence turn-off in Sco-Cen by
\citet{pec16} included seven additional stars of that kind.
All of those 21 stars have parallax and proper motion measurements 
from Hipparcos. Those data satisfy the kinematic criteria for membership 
from Section~\ref{sec:kin} for 11 of the 21 stars, although most of
the parallaxes have large errors by the standards of Gaia 
($\sigma_{\pi}>0.5$~mas). In addition, three of those 10 stars have
astrometry from Gaia DR2 that is inconsistent with membership.

The membership classifications in this work suggest that most previous
samples of candidate members of Sco-Cen have had substantial contamination
from field stars, which often was expected \citep{dez99}.
Even among the recent studies that utilize data from Gaia, there are
significant differences in distinguishing between field stars and members
of Sco-Cen and between members of the multiple populations in Sco-Cen,
which is a reflection of the variety of analysis methods and the overlapping
spatial locations and kinematics of those populations \citep{gal20b,luh20lu}.
To illustrate some of those differences in membership classifications,
I discuss two of the latest surveys in Sco-Cen that have used
data from Gaia EDR3, \citet{squ21} and \citet{gra21}.

\citet{squ21} selected candidate members of Upper Sco from Gaia EDR3 by
applying criteria that encompassed a concentration of stars within a space
defined by equatorial coordinates, parallax, and tangential velocities.
Thresholds on $\sigma_{\pi}/\pi$ and $G$ were also employed.
\citet{squ21} did not include criteria for selecting stars based
on evidence of youth, such as positions in CMDs (Section~\ref{sec:phot}).
When applying their selection criteria, I arrive at a sample that is slightly
larger than the size quoted in \citet{squ21} (2862 versus 2745).
They did not present a tabulation of their candidates, so the source of this
difference is unclear.
In my version of their sample, $\sim$69\% of the candidates satisfy my
criteria for membership in Upper Sco (Sections~\ref{sec:kin}, \ref{sec:phot}),
some of which have kinematics consistent with other Sco-Cen populations
as well. Among the remaining $\sim$31\% of candidates, roughly half
are rejected for membership in Sco-Cen by my criteria for kinematics or CMDs
and half are candidate members of other Sco-Cen populations, primarily UCL/LCC.
Through analysis of their candidates, \citet{squ21} concluded that Upper
Sco contains two distinct kinematic populations, one clustered and one
diffuse. They found that the diffuse population exhibited older
isochronal ages and a lower disk fraction than the clustered stars.
However, I find that their diffuse population is not part of Upper Sco,
and instead consists of a mixture of field stars and members of UCL/LCC,
which overlap with Upper Sco on the sky (Figure~\ref{fig:map}) and have
older ages and a lower disk fraction than Upper Sco \citep{luh20u,luh21}.

\begin{figure}
\epsscale{1}
\plotone{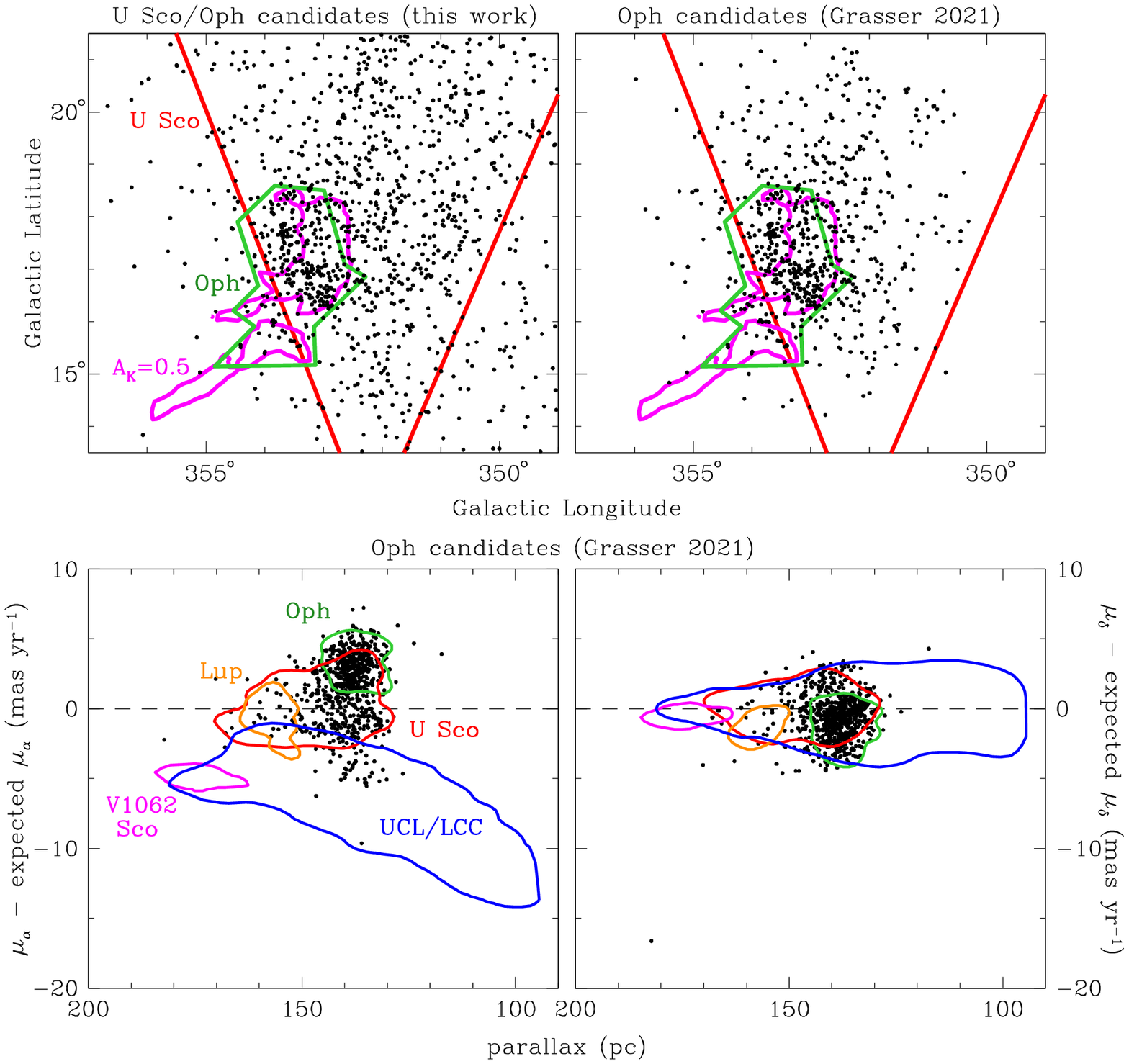}
\caption{
Top: Spatial distributions of stars from this work that have kinematics
consistent with membership in Upper Sco or Ophiuchus (left) and candidate
members of Ophiuchus from \citet{gra21} that have $\sigma_{\pi}<1$~mas (right). 
I have marked the boundary for Ophiuchus from \citet{esp18}
(green polygon), a field toward the center of Upper Sco defined by
\citet{luh20u} (red triangle), and contours for $A_K=0.5$ from the
extinction map of \citet{juv16} (magenta curves).
Bottom: Proper motion offsets versus parallactic distance for the candidates 
from \citet{gra21} are shown with the thresholds from Figures~\ref{fig:pp1}
and \ref{fig:pp2} that
are used in this work for selecting candidate members of populations in Sco-Cen.
}
\label{fig:mapp}
\end{figure}

To search for stars associated with the Ophiuchus clouds, \citet{gra21}
began by compiling young stars from previous studies that are located
within a large area extending beyond the clouds.
A subset of those stars with accurate kinematic data served as a training set 
for an algorithm that identified candidate members from Gaia EDR3 based on 
their proper motions and three-dimensional positions.
While examining their catalog of previous and new candidates, I found that
seven stars were listed twice, once with the Gaia DR2 designation and
once with the Gaia EDR3 designation. The numbers quoted in the following
discussion have been corrected for those duplicates.
Among the 150 stars in the training set from \citet{gra21},
my membership criteria (Sections~\ref{sec:kin}, \ref{sec:phot}) indicate
that 102 are candidate members of Ophiuchus (and other Sco-Cen populations
in some cases), 32 are candidates only for other Sco-Cen populations,
and 16 are not members of Sco-Cen.
In their total sample of 842 candidate members of Ophiuchus that have
$\sigma_{\pi}<1$~mas, I find that 412 and 570 have kinematics consistent with
Ophiuchus and Upper Sco, respectively (some are consistent with both),
while 117 are not members of any Sco-Cen population.
The parallactic distances and proper motion offsets for those 842 stars are 
shown with my membership criteria in Figure~\ref{fig:mapp}.

The presence of multiple Sco-Cen populations within the Ophiuchus sample
from \citet{gra21} is not surprising given that the training set was derived
from a compilation of previously identified young stars across a large area
that encompasses the Ophiuchus clouds as well as Upper Sco and UCL/LCC, as 
shown in Figures~\ref{fig:map} and \ref{fig:mapp}. In the latter, I have 
plotted maps of the Ophiuchus candidates from \citet{gra21} and the Ophiuchus
and Upper Sco candidates from this work. 
Indeed, \citet{gra21} found that their Ophiuchus candidates contained two
kinematic populations, and the one farther from the Ophiuchus
clouds was older on CMDs and exhibited a lower disk fraction, leading
them to conclude that it might be related to Upper Sco.
Finally, \citet{gra21} reported that their analysis had uncovered 190
new candidate members of Ophiuchus, but nearly all of those stars had
been previously identified as candidates for either Ophiuchus or Upper Sco,
most of which have been observed spectroscopically to verify their youth and 
have been examined for evidence of disks \citep{esp18,esp20,luh18,luh20u}.

Some of the previous candidates that are rejected in this work are only
modestly discrepant from the thresholds in distance and proper motion offsets
that I have adopted (Figures~\ref{fig:pp1} and \ref{fig:pp2}).
It is very likely that the populations in Sco-Cen contain at least a few
members that extend beyond those thresholds.
A small level of incompleteness is inevitable given the size and complexity
of Sco-Cen and is unimportant for most studies of its stellar population.
Members that are located in tails of the kinematic distributions 
are difficult to separate from field stars with only astrometry and photometry.
To identify such members, one would need to expand the kinematic thresholds
from Figures~\ref{fig:pp1} and \ref{fig:pp2}, obtain spectra for the additional 
candidates to confirm their youth and measure their radial velocities, and 
evaluate their membership in more detail with the resulting $UVW$ velocities.

\subsection{Comparison to X-ray Data from eROSITA}
\label{sec:xray}

Stars at ages of $\lesssim10$~Myr exhibit high ratios of X-ray to bolometric
luminosities \citep{fei99}.  As a result, X-ray emission has been used to 
identify candidate members of young clusters and associations 
\citep{pre01b,pre03,get02,get17,fei04,ste04,gud07,fei13}, including some of the
populations in Sco-Cen \citep{wal94,kra97,pre98,kun99,gro00,mam02,oza05}.
The extended Roentgen Survey with an Imaging Telescope Array (eROSITA) 
on board the Spectrum-Roentgen-Gamma mission \citep{pre21}
offers the best available combination of sensitivity, angular resolution, and 
spatial coverage for an X-ray survey of the entire Sco-Cen complex.
The instrument operates in soft X-rays (0.2--8~keV) and is imaging
the sky eight times during a period of four years.

\citet{sch21} have used eROSITA's first all-sky survey (eRASS1)
in conjunction with Gaia EDR3 to search for low-mass stars in Sco-Cen.
They identified matches between eRASS1 sources and stars from Gaia EDR3
that have parallactic distances of 60--200~pc, $\pi/\sigma_{\pi}>3$,
$G<17.5$, $G_{\rm BP}-G_{\rm RP}>1$ ($\gtrsim$G8), and 
locations within the Sco-Cen boundary from \citet{dez99}.
The resulting sample consisted of 6190 eRASS1 sources and their proposed
counterparts in EDR3. That catalog contains 40 instances in which 
two eRASS1 sources are matched to the same star from EDR3.
For each pair of possible matches, I have retained only the one that has the
higher probability of a correct match or (if the probabilities are equal)
the smaller separation.

\begin{figure}
\epsscale{1}
\plotone{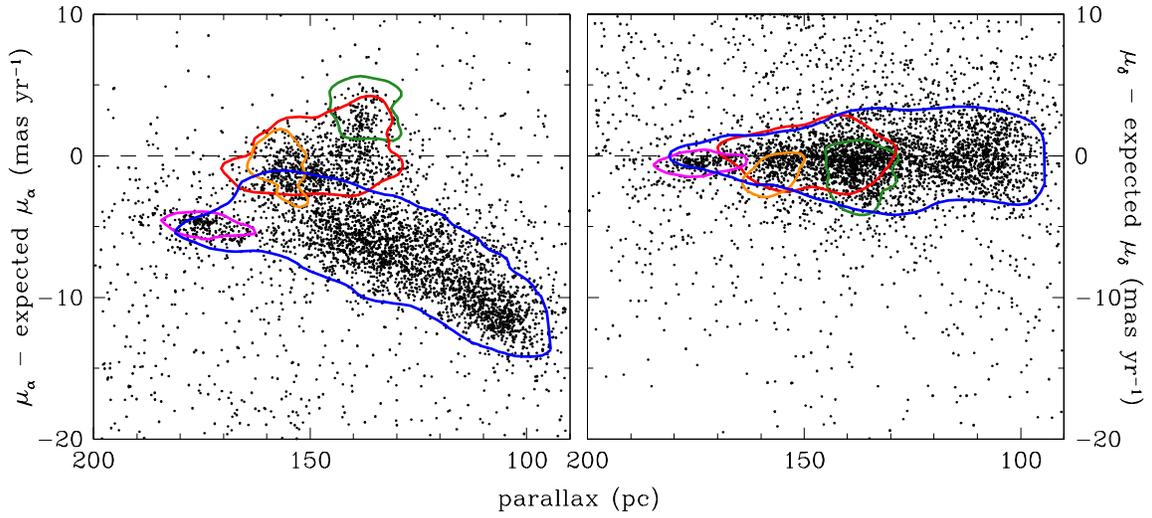}
\caption{
Proper motion offsets versus parallactic distance for eRASS1 sources toward 
Sco-Cen from \citet{sch21} are shown with the thresholds from 
Figures~\ref{fig:pp1} and \ref{fig:pp2} that are used in this work for 
selecting candidate members of populations in Sco-Cen.}
\label{fig:ppx}
\end{figure}

\citet{sch21} selected a subset of their eRASS1/EDR3 matches ($\sim90$\%) 
that might have appropriate ages for membership in Sco-Cen based on positions 
between model isochrones for 1 and 100~Myr in a Gaia CMD.
They found that most of the resulting candidates exhibited 
log~$L_{\rm X}/L_{\rm bol}$
near the saturation limit ($\sim-3$), which was indicative of youth, and that
some were well-clustered in parallax, tangential velocities, and
CMD ages while others formed a diffuse population in those parameters.
In Figure~\ref{fig:ppx}, I have plotted proper motion offsets and distances
for my reconstruction of that sample. My selection criteria from 
Figures~\ref{fig:pp1} and \ref{fig:pp2} are included for comparison.
The data show rich concentrations of stars and a sparse, widely-scattered
population, which correspond to the clustered and diffuse populations
described by \citet{sch21} and closely resemble the candidate young stars
that were presented in Figure~\ref{fig:pp1}. I find that $\sim54$\% of
the CMD-selected candidates from \citet{sch21} satisfy my kinematic and
CMD criteria for membership in any of the Sco-Cen populations.

\begin{figure}
\epsscale{1.2}
\plotone{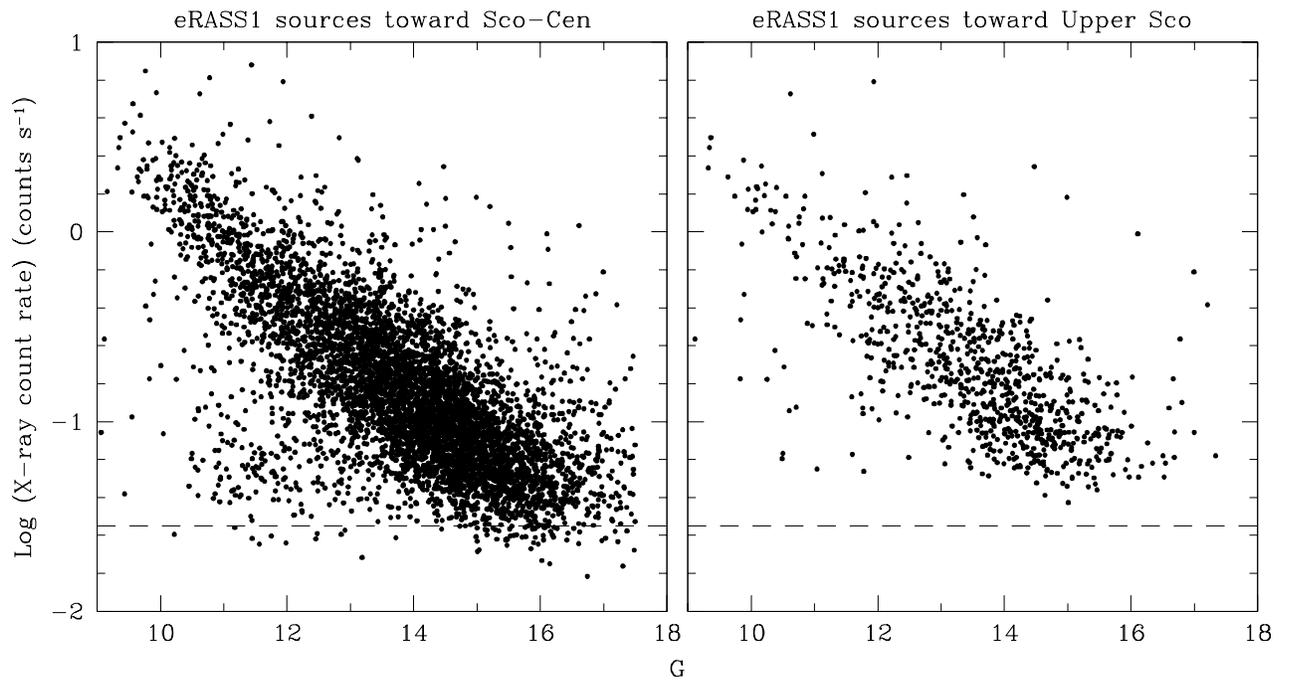}
\caption{
Count rate versus $G$ magnitude for eRASS1 sources from \citet{sch21}
toward Sco-Cen (left) and an area within Sco-Cen toward Upper Sco (right).
The typical flux limit adopted for Upper Sco in that study is indicated
(dashed line).
}
\label{fig:counts}
\end{figure}

\citet{luh20u} compiled a large sample of stars toward Upper Sco (excluding
Ophiuchus) that exhibit evidence of youth, primarily from spectroscopic 
diagnostics. The membership of each star was assessed with astrometry
from Gaia DR2 when available as well as proper motions and CMDs from
other sources \citep{luh18}.
\citet{sch21} estimated $L_{\rm X}/L_{\rm bol}$ for a subset of the
low-mass stars from \citet{luh20u} that had counterparts in the catalog
of eRASS1 sources toward Sco-Cen. Most of the estimates were near the
saturation limit, which was consistent with the previous evidence of youth
for those stars. \citet{sch21} also presented limits on $L_{\rm X}/L_{\rm bol}$ 
for Upper Sco candidates that lacked detections in eRASS1. 
At $G_{\rm BP}-G_{\rm RP}\lesssim3$, the limits were sufficiently low for
\citet{sch21} to question the youth, and hence the membership, of the stars.
To investigate that discrepancy, I have plotted in Figure~\ref{fig:counts}
count rate versus $G$ for all eRASS1 sources from \citet{sch21}
and for a subset toward Upper Sco. \citet{sch21} noted that the detection limit 
of eROSITA varies with position on the sky, and that the limits in Upper Sco 
are higher than in other parts of Sco-Cen, which is evident from 
Figure~\ref{fig:counts}. The plotted limits on $L_{\rm X}/L_{\rm bol}$ from
\citet{sch21} suggest a typical value of $\sim$0.028~counts~s$^{-1}$ for
the count rate limits that were adopted for Upper Sco in that study.
That value is well below the faintest eRASS1 fluxes toward Upper Sco,
as shown in Figure~\ref{fig:counts}, indicating that the limits were
underestimated. Based on the eRASS1 data toward Upper Sco, it appears that
the limit at which sources are reliably detected (i.e., the completeness
limit) in that areas of sky is $\sim$0.056~counts~s$^{-1}$.

\begin{figure}
\epsscale{1.2}
\plotone{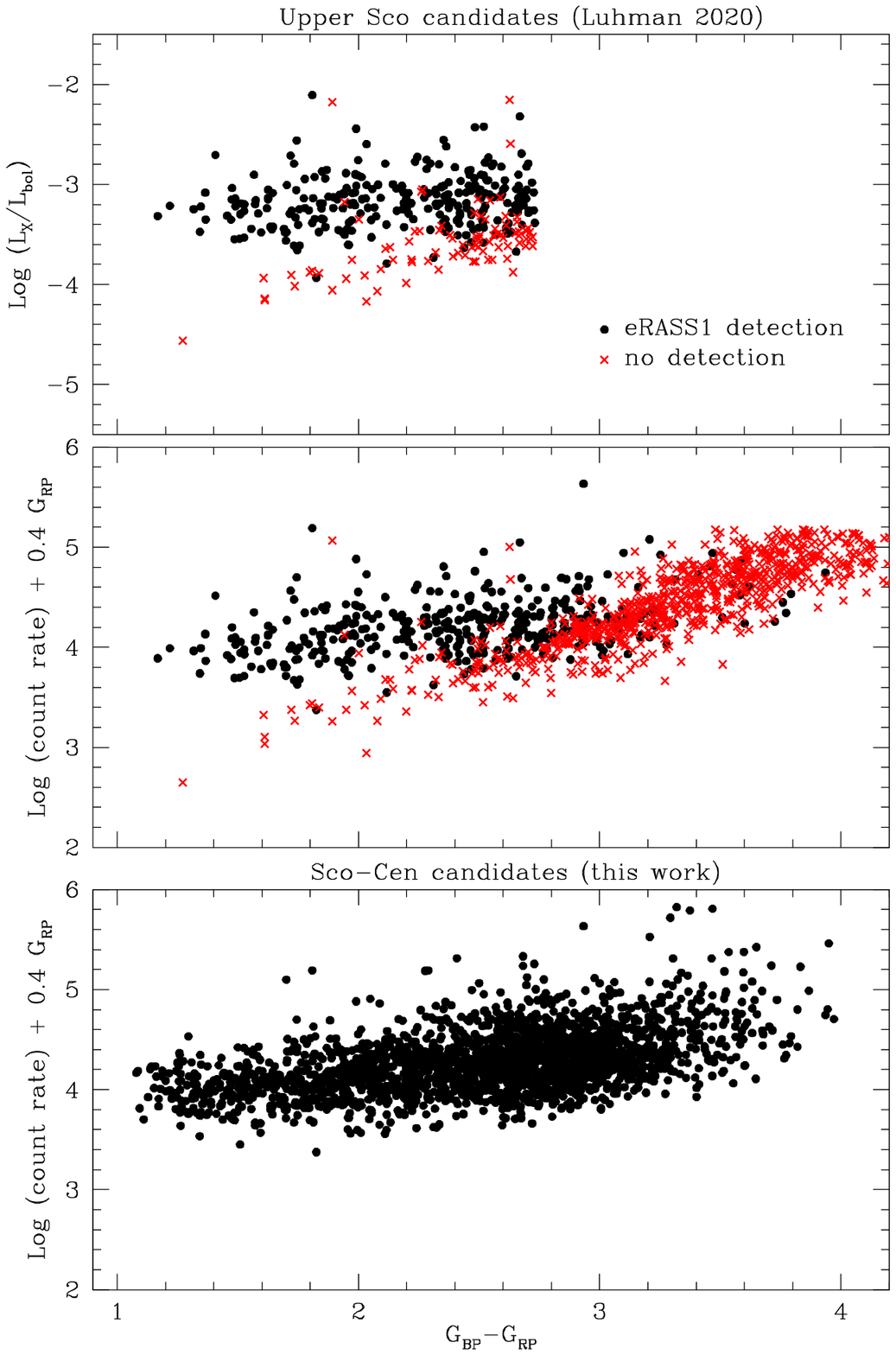}
\caption{
Top: $L_{\rm X}/L_{\rm bol}$ versus $G_{\rm BP}-G_{\rm RP}$ for candidate
members of Upper Sco from \citet{luh20u} based on eRASS1 data from 
\citet{sch21} and $G$-band bolometric corrections from \citet{and18}.
The latter are available for $T_{\rm eff}\gtrsim3300$~K 
($G_{\rm BP}-G_{\rm RP}\lesssim2.7$).
A limit of 0.056 counts~s$^{-1}$ has been assumed for the nondetections.
Middle: The candidates from \citet{luh20u} are plotted in terms of a
ratio of X-ray and $G_{\rm RP}$ fluxes (in arbitrary units), which 
does not require a bolometric correction.
Bottom: The same ratio is shown for candidate members of Sco-Cen
from this work (Section~\ref{sec:xray}).}
\label{fig:lx}
\end{figure}

In the top panel of Figure~\ref{fig:lx}, I have plotted 
$L_{\rm X}/L_{\rm bol}$ versus $G_{\rm BP}-G_{\rm RP}$ for Upper Sco
members from \citet{luh20u} that satisfy the criteria 
from \citet{sch21} (e.g., $G<17.5$, EDR3 parallaxes) and are not
close companions to other eRASS1 sources. For eRASS1 nondetections, 
a count rate limit of 0.056~counts~s$^{-1}$ has been adopted.
I have derived $L_{\rm X}$ and $L_{\rm bol}$ in the manner described
by \citet{sch21}. They employed $G$-band bolometric corrections
from \citet{and18}, which are applicable at $T_{\rm eff}=3300$--8000~K
($G_{\rm BP}-G_{\rm RP}\sim0.3$--2.7).
I also have calculated an observational parameter that is
analogous to $L_{\rm X}/L_{\rm bol}$ but does not rely on bolometric
corrections, namely a ratio of X-ray and $G_{\rm RP}$-band fluxes 
(in arbitrary units) that is defined as [log(count rate) + 0.4 $G_{\rm RP}$]. 
I have selected $G_{\rm RP}$ rather than $G$ to avoid contamination 
from accretion-related emission at UV wavelengths. This parameter is 
plotted versus $G_{\rm BP}-G_{\rm RP}$
in the middle panel of Figure~\ref{fig:lx}. Most of the eRASS1 detections
form a well-defined band that represents the saturation limit.
That band exhibits a small slope, which is a reflection of the
dependence of bolometric corrections on stellar temperature.
In the diagrams for both $L_{\rm X}/L_{\rm bol}$ and
[log(count rate) + 0.4 $G_{\rm RP}$], most of the nondetections are within
the band near the saturation limit when my estimate for the detection
limit is adopted, which is consistent with their other signatures of youth.
Only a small number of nondetections, those at
$G_{\rm BP}-G_{\rm RP}\lesssim2.4$, appear significantly below the
saturation limit. All of those stars exhibit Li absorption that is strong
enough to indicate youth, and roughly half have additional evidence
of youth in the form of IR excesses.
That disk fraction is higher than the value measured for members of 
Upper Sco in the same range spectral types \citep{luh20u,luh21},
and disk-bearing stars tend to be fainter in X-rays \citep{pre05,tel07},
so some of the X-ray nondetections may be related to the presence of disks.
Meanwhile, it appears that solar-mass stars can begin to experience a
decay in $L_{\rm X}$ and $L_{\rm X}/L_{\rm bol}$ by the time they reach
the age of Upper Sco \citep[][K. Getman, in preparation]{gre16},
which might account for some of the nondetections.
It is also possible that the nondetections have normal X-ray
luminosities for young stars, but they were not measured in eRASS1 due
to issues related to the instrument or data processing.
Finally, I note that the data in Figure~\ref{fig:lx} indicate that eRASS1
has a high level of completeness among Upper Sco members for 
$G_{\rm BP}-G_{\rm RP}\lesssim2.5$ ($\lesssim$M3). 

I can use the eRASS1 data to assess the youth of the low-mass stars that
I have identified as candidate members of Sco-Cen, most of which lack
spectroscopic observations.
In Section~\ref{sec:field}, I described a sample of 7272 Sco-Cen candidates
in which $G_{\rm BP}$ and $G_{\rm RP}$ are available and field star
contamination is minimized. For the 2289 candidates from that sample
that have detections in eRASS1 from \citet{sch21}, I have plotted 
[log(count rate) + 0.4 $G_{\rm RP}$] versus $G_{\rm BP}-G_{\rm RP}$
in the bottom panel of Figure~\ref{fig:lx}.
Virtually all of the candidates are found in a well-defined band
that matches the band observed among confirmed young stars in Upper Sco,
which is consistent with the youth implied by the CMDs. The thousands of 
candidates that lack eRASS1 detections are absent from Figure~\ref{fig:lx} 
since their individual count rate limits are not available, but if
the typical limits implied by Figure~\ref{fig:counts} were adopted,
most of the nondetections would fall near the band exhibited by the
detections, and thus would be consistent with youth.

\section{Properties of the Stellar Populations in Sco-Cen}
\label{sec:pop}

\subsection{Adopted Samples}
\label{sec:clean}

In Sections~\ref{sec:kin} and \ref{sec:phot}, I identified 
candidate members of populations in Sco-Cen based on their clustering
in proper motion offsets and distance and their positions in CMDs.
One can further refine the sample for a given population
by considering only the field on the sky where the candidates are
concentrated (i.e., requiring clustering in celestial coordinates).
By doing so, the probabilities of membership for the candidates should
be maximized and the contamination from field stars and other Sco-Cen
populations should be minimized.
Therefore, when characterizing the stellar populations in Sco-Cen
using the candidates in Table~\ref{tab:cand}, I consider the
following samples of candidates with RUWE$<$1.6, $\sigma_{\pi}<0.5$~mas, and 
$\sigma_{BP}<0.1/\sigma_{RP}<0.1$ or $\sigma_G<0.1/\sigma_{RP}<0.1$:
1) Ophiuchus kinematics and a location
within the boundary of Ophiuchus from \citet{esp18};
2) Upper Sco kinematics and a location outside of
Ophiuchus and within the triangular field from \citet{luh20u}; 
3) Lupus kinematics and a location within the fields toward clouds 1--4
from \citet{luh20lu}; 
4) V1062~Sco kinematics and a location within a radius
of $2\arcdeg$ from the center of that group; 
and 5) UCL/LCC kinematics and a location within the boundary from 
\citet{dez99} and not within the fields for the other four samples.
Thus, I do not attempt to study more diffuse populations
outside of these fields (see Figure~\ref{fig:map}) with the exception of
UCL/LCC. As a reminder, the kinematic criteria for the populations are shown in
Figures~\ref{fig:pp1} and \ref{fig:pp2} and a flag in Table~\ref{tab:cand}
indicates the populations whose criteria are satisfied by each candidate.

\subsection{Spectral Types and Extinctions}
\label{sec:spt}

Some of the analysis of the candidates in Sco-Cen requires estimates of
spectral types and extinctions.  The derivation of those estimates in this
section makes use of the typical values of the intrinsic colors of young
stellar photospheres as a function of spectral type.  \citet{luh20u} estimated
those colors for a selection of standard optical and IR bands (including those 
from Gaia DR2) using data for known members of several nearby star-forming 
regions and young associations ($\lesssim20$~Myr).
I have updated that analysis to include the bands from Gaia EDR3, which
have a slightly different photometric system than the bands from DR2
\citep{rie21}, and to make use of the candidate members of Sco-Cen in
Table~\ref{tab:cand} that have measured spectral types and low extinction.
The new estimates of the intrinsic colors of young stars and brown dwarfs
are presented in Table~\ref{tab:intrinsic}.

\begin{figure}
\epsscale{1.4}
\plotone{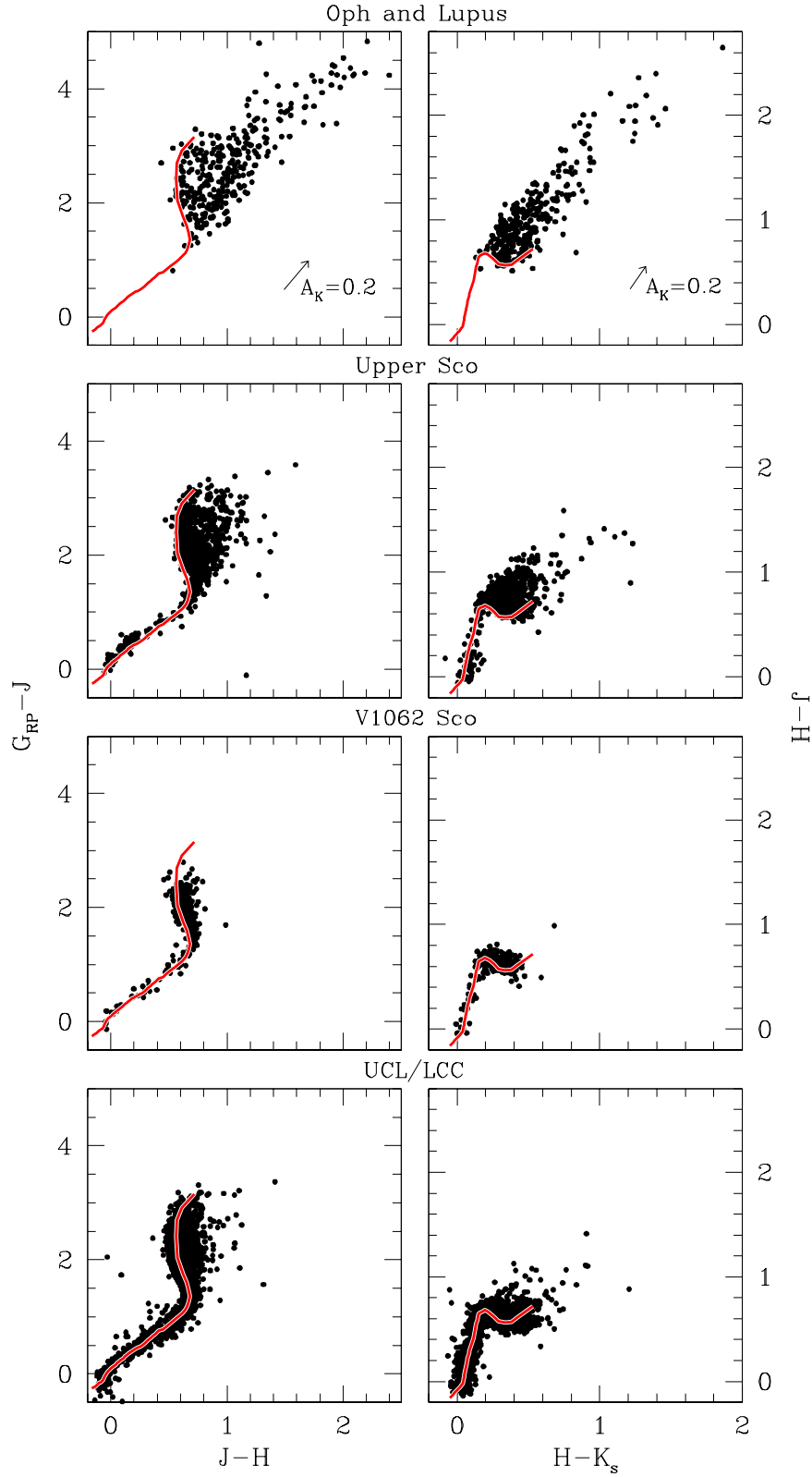}
\caption{
$G_{\rm RP}-J$ versus $J-H$ and $J-H$ versus $H-K_s$ for candidate members
of populations in Sco-Cen as defined in Section~\ref{sec:clean}.
The intrinsic colors of young stars from B0--M9 are indicated
(red lines, Table~\ref{tab:intrinsic}).
}
\label{fig:cc}
\end{figure}

To illustrate the levels of extinction in the Sco-Cen populations, I have
plotted color-color diagrams with $G_{\rm RP}-J$, $J-H$, and $H-K_s$
in Figure~\ref{fig:cc} for the samples defined in Section~\ref{sec:clean}.
These bands were selected to minimize short-wavelength emission related to
accretion and long-wavelength emission from cool dust in systems that have 
circumstellar disks. Reddening is largest for Ophiuchus and Lupus, remains 
noticeable for Upper Sco, and is negligible for V1062~Sco and UCL/LCC 
($A_K<0.05$). That trend is correlated with with the relative ages of the 
populations \citep[][references therein]{luh20u,luh20lu}.
The extinctions among members of Ophiuchus extend to much higher values
than implied by Figure~\ref{fig:cc} since many of the previously
proposed members are too obscured for detections with an optical telescope.
Only a few likely members of Lupus are absent from Figure~\ref{fig:cc}
due to high extinction. Most of the outliers in the color-color diagrams are
secondaries in multiple systems for which the marginally-resolved
photometry from 2MASS is likely unreliable and stars that experienced
variability between the 2MASS and Gaia measurements.

For the candidates that have spectral classifications,
I have derived extinctions from color excesses in $G_{\rm RP}-J$, $J-H$,
$G-G_{\rm RP}$, or $G_{\rm BP}-G_{\rm RP}$ (in order of preference)
relative to the intrinsic colors of young stars at a given spectral type.
Extinctions in $K_s$ are calculated from the color excesses using
relations from \citet{ind05} and \citet{sch16} and the following relations
that are derived from reddened members of Upper Sco and Ophiuchus: 
$E(G_{\rm RP}-J)/E(J-H)\approx2.4$, $E(G-G_{\rm RP})/E(J-H)\approx0.6$,
and $E(G_{\rm BP}-G_{\rm RP})/E(J-H)\approx3$.
A small number of candidates are close companions that have photometry
in only a single band ($G$), so they lack the color measurements needed
for extinction estimates. Objects of this kind are omitted from the 
samples defined in Section~\ref{sec:clean} by the requirement for photometry
in $G_{\rm BP}/G_{\rm RP}$ or $G/G_{\rm RP}$.

I have estimated spectral types and extinctions for the candidates
that lack spectroscopy by dereddening their observed colors to the sequences
of intrinsic colors of young stars in diagrams of $G_{\rm RP}-J$ versus $J-H$
and $J-H$ versus $H-K_s$ (Figure~\ref{fig:cc}).
In those diagrams, the sequences of intrinsic colors at $\leq$M0 overlap
with the reddened colors of stars at earlier types. In other words,
most stars redder than $G_{\rm RP}-J\sim1.3$ and $J-H\sim0.7$ can be
dereddened to either of two points on the sequence, one at early types
and one at late types, leading to a degeneracy in the estimates of
spectral type and extinction. This degeneracy can be largely broken by
calculating the dereddened absolute magnitudes for each of the two
possible combinations of spectral type and extinction and checking which
spectral type is better supported by the absolute magnitudes given the
CMD for the population. For instance, among the Sco-Cen candidates that
are red enough to suffer from this degeneracy (and lack spectral types), none
would have dereddened photometry that would be bright enough for an early-type 
member, so they are dereddened to $\geq$M0 portions of the sequences.
If a candidate has a measurement of a Gaia color but lacks IR photometry,
which is the case for some companions, the spectral type is estimated from
that color with the assumption that extinction is absent.

\subsection{Initial Mass Function}

\begin{figure}
\epsscale{1.2}
\plotone{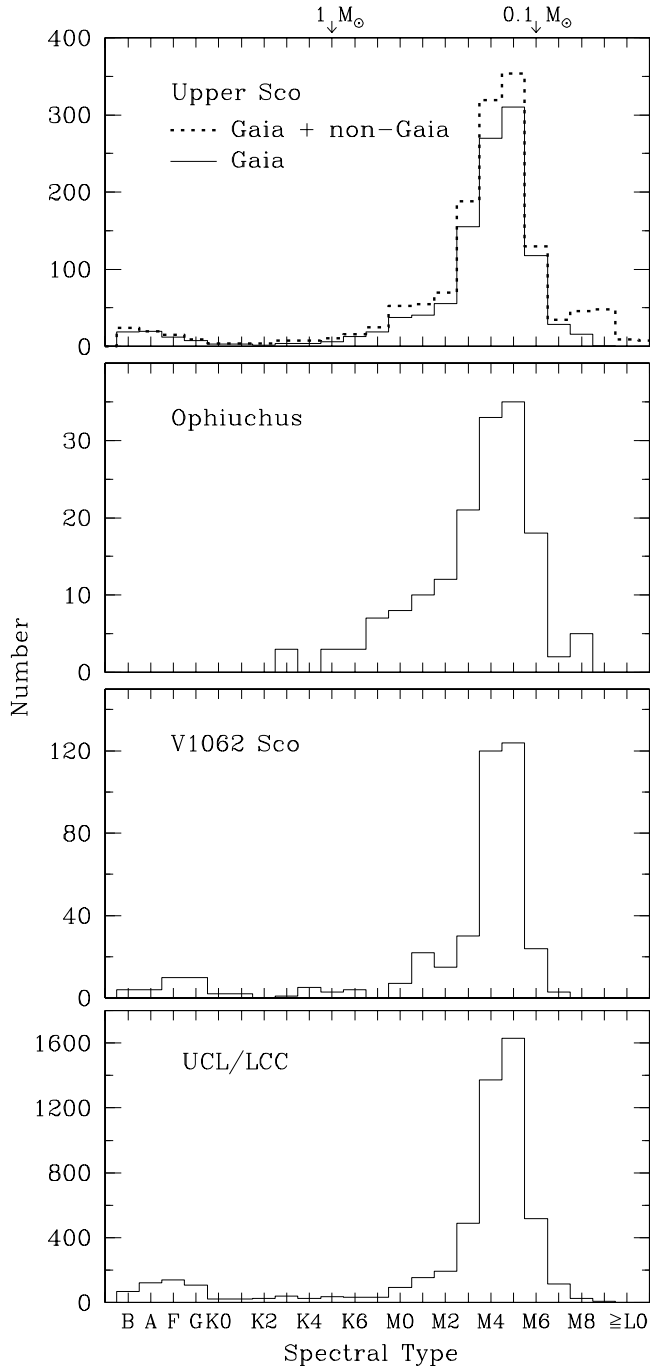}
\caption{
Histograms of spectral types for candidate members of Upper Sco,
Ophiuchus ($A_K<0.5$), V1062~Sco, and UCL/LCC that satisfy the criteria
in Section~\ref{sec:clean} (solid histograms).
For stars that lack spectroscopy, spectral types have been
estimated from photometry (Figure~\ref{fig:cc}, Section~\ref{sec:spt}).
A second histogram is shown for Upper Sco that includes additional young
stars that are absent from the Gaia sample and have been identified in
spectroscopic surveys \citep[dotted histogram,][references therein]{luh20u}.
The arrows mark the spectral types that correspond to masses of 0.1
and 1~$M_\odot$ for ages of 10~Myr according to evolutionary models
\citep[e.g.,][]{bar98}.
}
\label{fig:histo}
\end{figure}

The histogram of spectral types in a young stellar population can serve
as an observational proxy for its IMF.
In Figure~\ref{fig:histo}, I have plotted such histograms for the samples of
candidates in Upper Sco, Ophiuchus, V1062~Sco, and UCL/LCC defined in
Section~\ref{sec:clean}. The histograms employ spectroscopic classifications
from previous studies when available (Table~\ref{tab:cand}) and otherwise use
the photometric estimates from the previous section.
The Lupus sample has been omitted since its IMF was characterized with
Gaia DR2 \citep{gal20b,luh20lu} and EDR3 produces similar results.
Because the Ophiuchus candidates have the largest range of extinctions among 
the Sco-Cen populations and brighter, more massive stars can be detected
at higher extinctions, those candidates are likely to be biased in favor of
earlier spectral types. To reduce such biases, only candidates with $A_K<0.5$
are included in the histogram for Ophiuchus.
Since Ophiuchus overlaps kinematically and spatially with Upper Sco 
(Figure~\ref{fig:pp1}), the Ophiuchus sample likely includes at least a few 
members of Upper Sco.

In Figure~\ref{fig:histo}, the histograms for Upper Sco, Ophiuchus, V1062~Sco, 
and UCL/LCC exhibit peaks near M4--M5, which corresponds to a mass of 
$\sim0.15$--0.2~$M_\odot$ for ages of $\lesssim20$~Myr \citep{bar98,bar15}. 
The same is true for Lupus as well \citep{gal20b,luh20lu}.  
Thus, all of the populations in Sco-Cen share
similar IMFs in terms of their characteristic masses. The histograms
of spectral types in Sco-Cen are also similar to those measured in other
nearby star-forming regions like Perseus and Taurus \citep{luh16,esp19}.

Some of the previous surveys for members of Upper Sco 
have used imaging and spectroscopy that reach lower masses 
($\sim0.01$~$M_\odot$) than the data from Gaia 
\citep[][references therein]{luh18}, and thus can be used to constrain
the completeness of the sample of candidates in Upper Sco from this work.
Based on data from a variety of imaging surveys, previous studies
have spectroscopically identified nearly all young stars earlier than $\sim$M9
that have CMD positions and proper motions consistent with Upper Sco
membership and that are located outside of the Ophiuchus boundary
from \citet{esp18} and within the triangular field in the center
of Upper Sco in Figure~\ref{fig:map} \citep{luh20u}.
I have compiled all known young objects of that kind that are located within
that field and that are absent from my Gaia sample of Upper Sco candidates
(e.g., they lack Gaia parallaxes with $\sigma_{\pi}<0.5$~mas) and I have
combined them with the Gaia sample in a histogram shown in 
Figure~\ref{fig:histo}. The comparison of that histogram to the histogram
of Gaia candidates suggests that the completeness of the latter sample
is $\gtrsim$85--90\% at spectral types of M0--M7 and rapidly decreases
at later types. That value is a lower limit since some of the non-Gaia objects
may not be members of Upper Sco (e.g., members of UCL/LCC). 
These results are consistent with the completeness estimated in 
Section~\ref{sec:completeness} for EDR3 sources satisfying the criteria applied 
to the samples of Sco-Cen candidates in Figure~\ref{fig:histo}.
The sample of candidates in UCL/LCC likely becomes increasingly incomplete
later than M7 ($\lesssim0.06$~$M_\odot$) like Upper Sco since its older age
is roughly canceled by the smaller distances of most of its members. 
That limit in the V1062~Sco group likely occurs at a slightly earlier type
of $\sim$M6 ($\sim0.1$~$M_\odot$) given that it is older and more distant
than Upper Sco.

Table~\ref{tab:cand} contains 8000 candidate members of the Sco-Cen complex
that have RUWE$<$1.6, $\sigma_{\pi}<0.5$~mas, $\sigma_{BP}<0.1/\sigma_{RP}<0.1$ 
or $\sigma_G<0.1/\sigma_{RP}<0.1$,
and locations within the Sco-Cen boundary from \citet{dez99}. 
Among those candidates, $\sim$95\% have spectral type estimates 
of $\leq$M6, which correspond to stellar masses according to 
evolutionary models \citep{bar98}. If this sample of stellar candidates is 
$\sim$90\% complete, then Sco-Cen would contain a total of $\sim$8300 stars. 
If the ratio of stars to brown dwarfs at 0.01--0.08~$M_\odot$ in Upper Sco
and other nearby star-forming regions
\citep[$N_{star}/N_{BD}\sim7$,][]{luh16,luh20u,esp19} applies to all
populations in Sco-Cen, then the total number of brown dwarfs in that
mass range would be $\sim$1200. Thus, my survey suggests that the Sco-Cen
complex (as defined in this work) likely contains nearly 10,000 stars
and brown dwarfs.
Previous studies have arrived at comparable estimates for the size 
of the stellar population in Sco-Cen \citep[e.g.,][]{mam02,pm08,dam19,sch21}.

\subsection{Stellar Ages}
\label{sec:ages}

The ages of the stellar populations in Sco-Cen have been previously constrained 
with several methods \citep{deg89,mam02,pre02,pm08,pec12,son12,fei16,pec16}.
The astrometry and photometry from Gaia facilitate such work
by providing reliable membership samples and accurate measurements of the
sequences formed by low-mass stars in the Hertzsprung-Russell (H-R) diagram.
Data from Gaia DR2 have been used to construct H-R diagrams in Sco-Cen that
employ Gaia magnitudes and colors 
\citep{gol18,dam19,esp20,luh20u,luh20lu,ker21}, $M_K$ and spectral types 
\citep{esp20}, and estimates of temperature and luminosity from fits
to spectral energy distributions \citep{gal20b}.
Based on the age estimated for the $\beta$~Pic association from
its lithium depletion boundary \citep[21--23~Myr,][]{bin16} and the
change in luminosity with age predicted by evolutionary models
\citep{bar15,cho16,dot16,fei16}, the sequences of low-mass stars
in Sco-Cen relative to the $\beta$~Pic sequence have implied ages of 2--6~Myr
for groups in Ophiuchus, $\sim$6~Myr for Lupus, $\sim$11~Myr for Upper Sco, and
$\sim$20~Myr for V1062~Sco and UCL/LCC \citep{esp20,luh20u,luh20lu}.

\begin{figure}
\epsscale{1.2}
\plotone{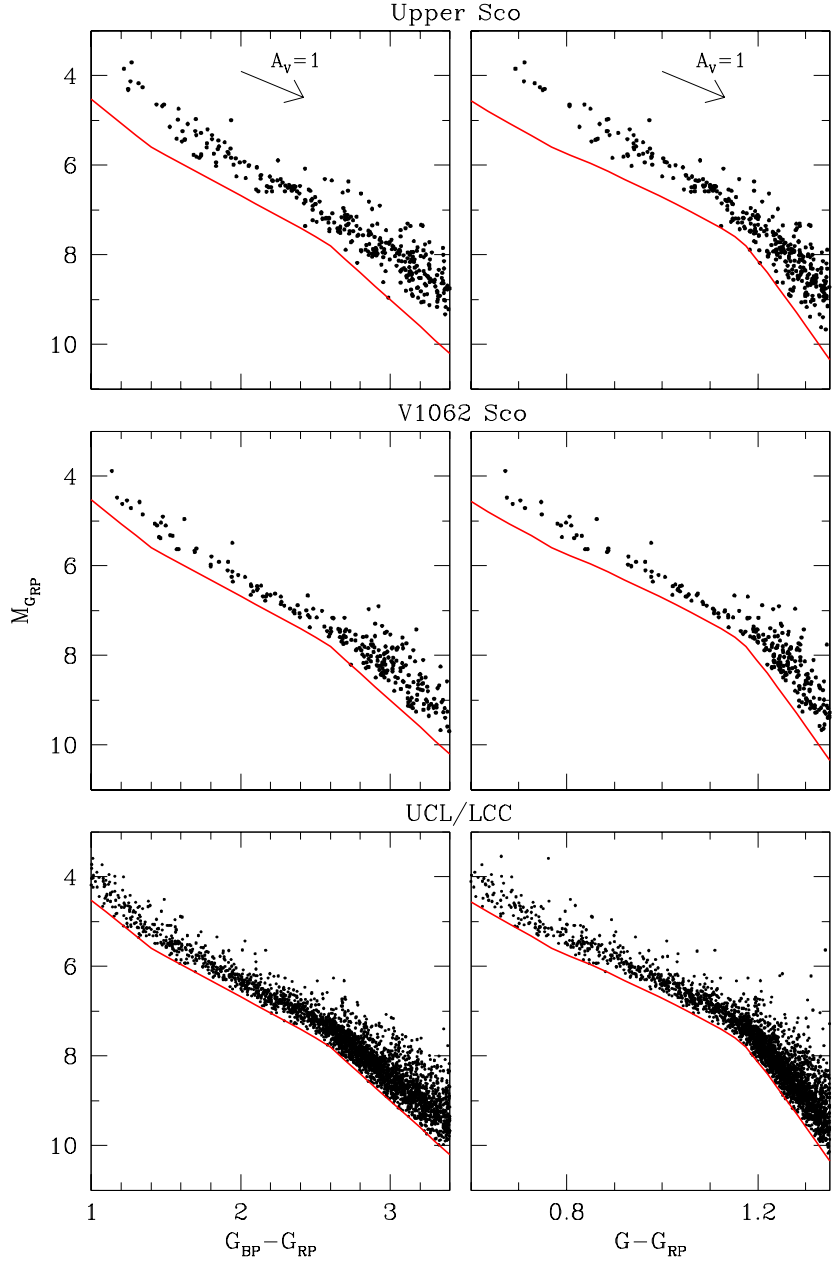}
\caption{
$M_{G_{\rm RP}}$ versus $G_{\rm BP}-G_{\rm RP}$ and $G-G_{\rm RP}$ for 
candidate late-type ($\sim$K0--M5) members of Upper Sco, V1062~Sco, and UCL/LCC
that have $\sigma_{\pi}<0.1$~mas and $A_K<0.1$ and do not have full disks.
The boundaries from Figure~\ref{fig:cmd} that were used for selecting candidates
have been included (red lines).
}
\label{fig:cmd3}
\end{figure}

The previous age estimates based on Gaia DR2 are not significantly affected by
updating them to include EDR3 and the new membership assignments in this work.
To illustrate the relative ages for the older populations in Sco-Cen,
I show in Figure~\ref{fig:cmd3} the CMDs for candidate low-mass stars
in Upper Sco, V1062~Sco, and UCL/LCC that satisfy the criteria
in Section~\ref{sec:clean} as well as $\sigma_{\pi}<0.1$~mas, $A_K<0.1$, and 
the absence of full disks \citep{luh21}. The data in those CMDs have been
corrected for the extinctions estimated in Section~\ref{sec:spt}.
The ages implied by the CMDs are sensitive to extinction errors and
the adopted reddening law, which is why only stars with low extinction have
been included in the CMDs. Stars with full disks have
been excluded since accretion-related emission at shorter optical
wavelengths can produce systematic errors in the ages inferred from CMDs.
As done in \citet{luh20u}, I have calculated the offset in $M_{G_{\rm RP}}$ 
for each star from a fit to the median of the sequence for UCL/LCC. 
For candidates between $G_{\rm BP}-G_{\rm RP}=1.4$--2.8
($\sim$0.2--1~$M_\odot$, K5--M4), the medians of the offsets for
Upper Sco and V1062~Sco are brighter than the median offset for
UCL/LCC by $0.34\pm0.02$ and $0.02\pm0.03$~mag, respectively, which
are consistent with the results from \citet{luh20u}.

Since UCL/LCC is much more extended than the other populations in Sco-Cen,
it is useful to check whether the ages of its members vary with location.
I have calculated the median $M_{G_{\rm RP}}$ offset for each UCL/LCC
candidate from Figure~\ref{fig:cmd3} and its six nearest neighbors in $XYZ$.
The resulting median offsets are plotted versus Galactic longitude and on a map 
of Galactic coordinates in Figure~\ref{fig:amap}.
The small number of candidates with $b\lesssim0\arcdeg$ are systematically
brighter by $\sim0.25$~mag, which would suggest an age of $\sim14$~Myr if the
remainder of UCL/LCC has an age of 20~Myr. The younger ages of those stars have
been noticed in previous analysis of Gaia data \citep{gol18,ker21}. 
The opposite end of UCL/LCC at $l\gtrsim340\arcdeg$, which overlaps with
Upper Sco, also may have a slight excess of younger stars relative to the
bulk of the population.
The remaining candidates in UCL/LCC do not show systematic variations in age
with position, and the median age is fairly constant across the length of the
population in Galactic longitude.
Meanwhile, \citet{pec16} found that their sample of candidate members of 
Sco-Cen exhibited variations of average age with celestial coordinates.
Those gradients were likely a reflection of the overlap among the Sco-Cen
populations on the sky (Figure~\ref{fig:map}),  which were difficult to
separate prior to the availability of the precise astrometry from Gaia.
Field star contamination may have contributed to the gradients as
well (Table~\ref{tab:prev}).

\subsection{Radial Velocities and $UVW$ Velocities}
\label{sec:uvw}

Measurements of radial velocities with errors less than 4~km~s$^{-1}$
are available for 1740 candidates from Table~\ref{tab:cand}.
As done in \citet{luh20lu}, I have adopted errors of 0.4 and 1~km~s$^{-1}$
for velocities from \citet{tor06} and \citet{wic99} that lacked
reported errors, respectively. I have combined the radial velocities
with proper motions from EDR3 and parallactic distances based on EDR3
parallaxes \citep{bai21} to compute $UVW$ space velocities \citep{joh87},
which are included in Table~\ref{tab:cand}. The velocity errors were
estimated in the manner described by \citet{luh20u}.

Before examining the $UVW$ velocities, I plot in the top row of
Figure~\ref{fig:uvw} the $XYZ$ positions in Galactic Cartesian coordinates
for all candidates from Table~\ref{tab:cand} that have RUWE$<$1.6,
$\sigma_{\pi}<0.5$~mas, $\sigma_{BP}<0.1/\sigma_{RP}<0.1$ or
$\sigma_G<0.1/\sigma_{RP}<0.1$,
and locations within the Sco-Cen boundary from \citet{dez99}.
I have included density contours that encompass most of the candidates
that are within the fields in Upper Sco, Ophiuchus, Lupus, and V1062~Sco
that are marked in the map of Galactic coordinates in Figure~\ref{fig:map}.
According to those diagrams, the dimensions of the Sco-Cen complex range
from 40--120~pc, which is consistent with results from previous Gaia
surveys \citep{wri18,dam19}.

In the middle row of Figure~\ref{fig:uvw}, I have plotted $XYZ$ 
for the candidates that have measurements of radial velocities (and hence
$UVW$) and that are in the samples defined in Section~\ref{sec:clean}.
The colors of the points are assigned according to the samples to which
they belong. For Figure~\ref{fig:uvw}, candidate members of UCL/LCC are
divided into two samples using the boundary between UCL and LCC 
($l=312\arcdeg$) defined by \citet{dez99}.
A comparison of the top and middle rows in Figure~\ref{fig:uvw} indicates
that the available radial velocities provide a good sampling of the 
various populations in Sco-Cen. The medians and standard deviations of 
the $UVW$ velocities for these samples are presented in Table~\ref{tab:uvw}.

The precise astrometry from Gaia has been widely utilized for measuring the
internal kinematics of OB associations and rich star-forming clusters
\citep{gol18,kou18,war18,wri18,can19b,kuh19,wri19,mel20b,zar19,swi21}.
Some studies have found evidence of expansion while others have not
\citep{wri20}. Both results have been reported for Sco-Cen \citep{gol18,wri18}.
Expansion along a given axis is manifested by a positive correlation between
the velocity and position along that axis.
To check for evidence of expansion in Sco-Cen, I have plotted $U$, $V$, and $W$ 
versus $X$, $Y$, and $Z$, respectively, in the bottom row of
Figure~\ref{fig:uvw}.
The candidates in UCL/LCC exhibit clear correlations along each axis.
\citet{gol18} detected similar evidence of expansion in LCC using Gaia DR2.
Upper Sco also shows correlations along $U$ and $V$.
Correlations are not apparent within the other samples, which are much more
compact than UCL/LCC and Upper Sco, and the bulk motions of those samples do 
not show a pattern of expansion. For instance, according to
Figure~\ref{fig:uvw}, Upper Sco and Ophiuchus are moving away from UCL/LCC in
$Y$ but not in $X$ and $Z$.  The coherent pattern of expansion among UCL and 
LCC candidates combined with their continuous spatial distribution
(Figure~\ref{fig:map}) and relatively uniform median age with location 
(Figure~\ref{fig:amap}) indicate that UCL and LCC comprise a single stellar
population.

\section{Conclusions}

\begin{figure}
\epsscale{1.2}
\plotone{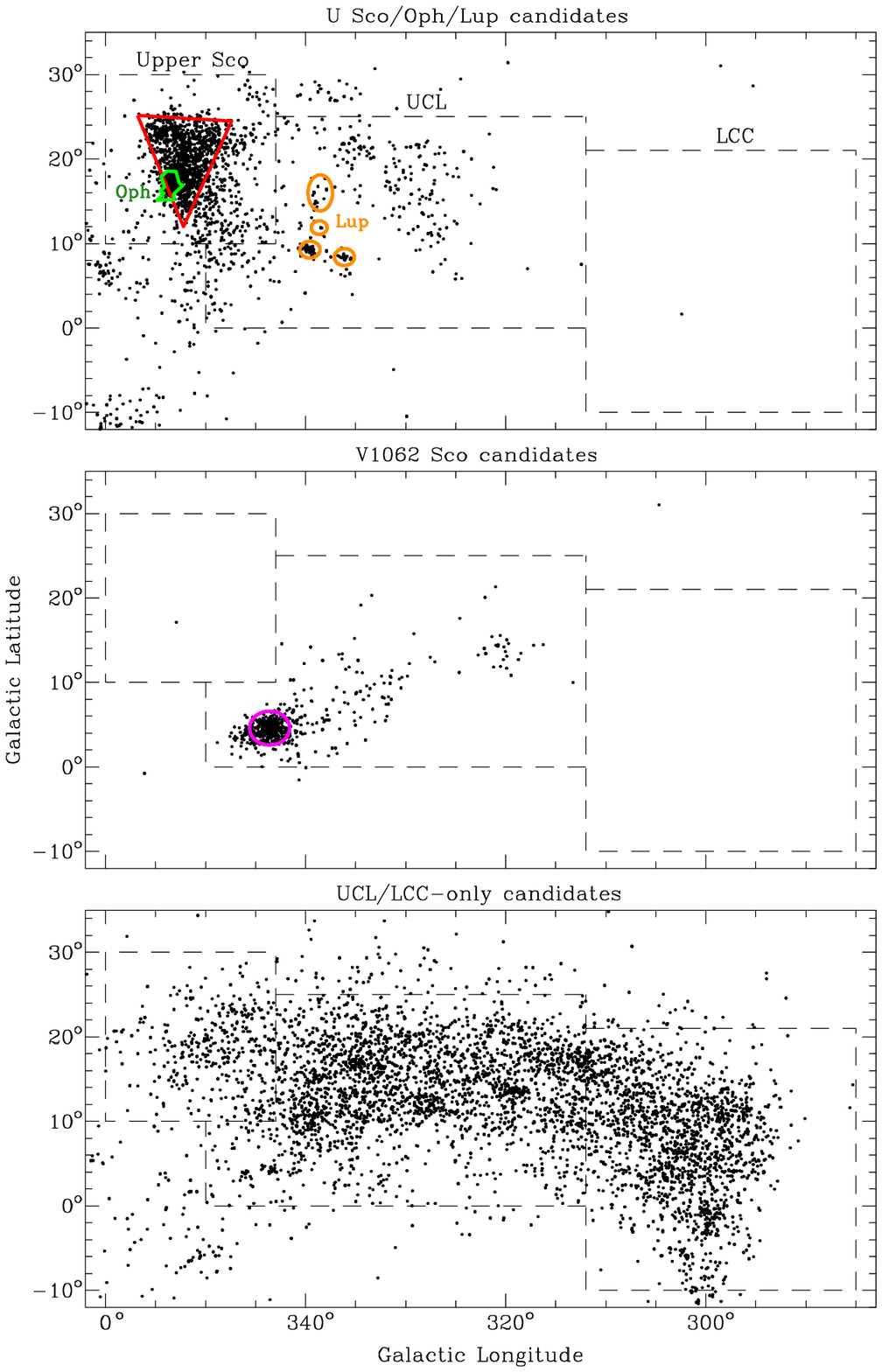}
\caption{
Spatial distribution of candidate members of populations in Sco-Cen based
on the kinematic and photometric criteria in 
Figures~\ref{fig:pp1}--\ref{fig:cmd}.  
I have marked the boundaries for Upper Sco, UCL, and LCC from \citet{dez99}
(dashed lines), the boundary for Ophiuchus from \citet{esp18} (green polygon),
a field toward the center of Upper Sco defined by \citet{luh20u} (red triangle),
the fields encompassing Lupus clouds 1--4 from \citet{luh20lu}
(orange ellipses), and a $2\arcdeg$ radius field toward V1062~Sco
(magenta ellipse).
}
\label{fig:map}
\end{figure}

\begin{figure}
\epsscale{1.4}
\plotone{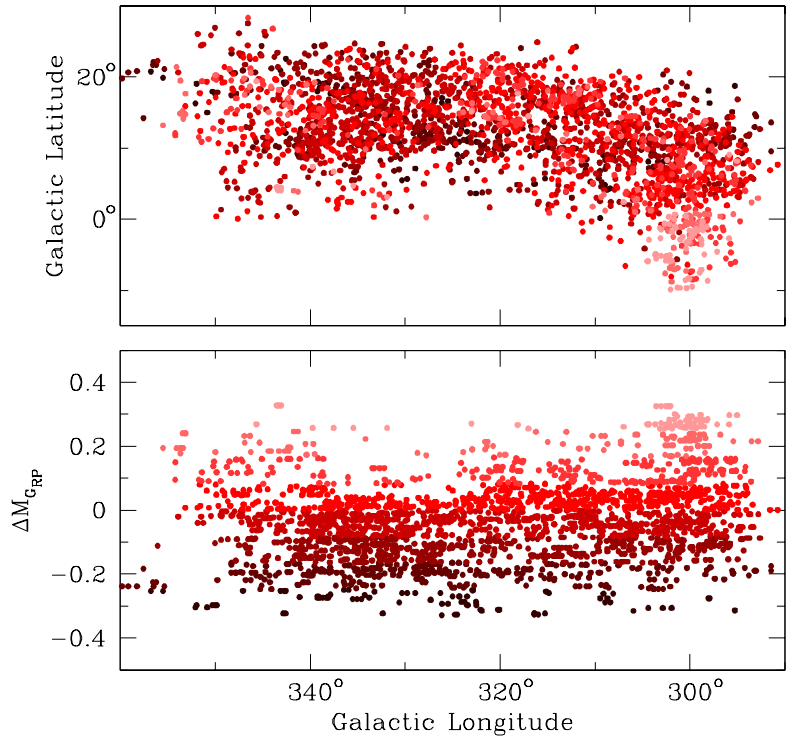}
\caption{
Median offsets in $M_{G_{\rm RP}}$ from the median sequence of UCL/LCC 
for each candidate member of UCL/LCC from Figure~\ref{fig:cmd3} and its
six closest neighbors in $XYZ$. The offsets are defined so that positive
values correspond to younger ages. The colors of the points correspond to the
values of the $M_{G_{\rm RP}}$ offsets.
}
\label{fig:amap}
\end{figure}

\begin{figure}
\epsscale{1.1}
\plotone{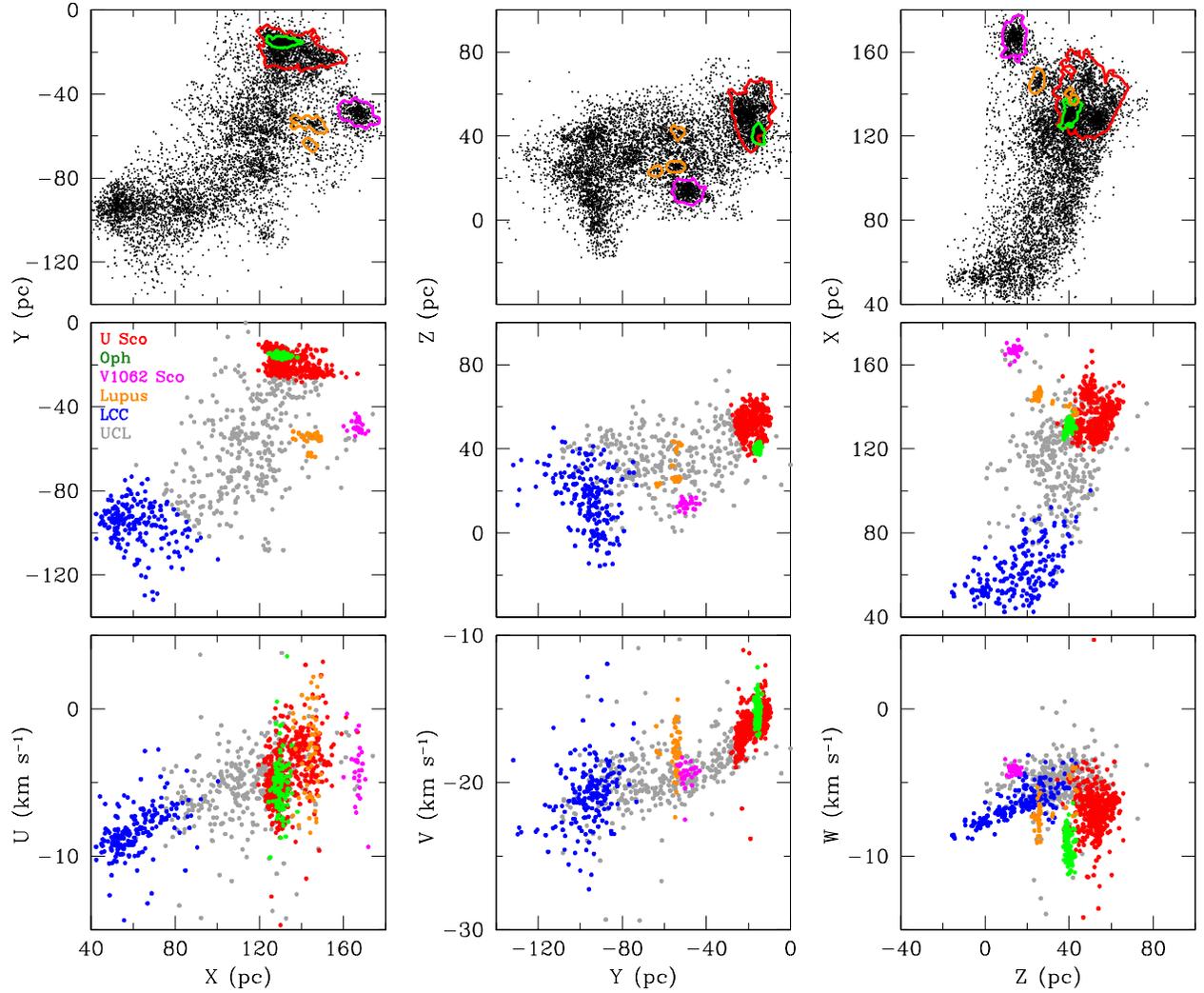}
\caption{
Galactic Cartesian coordinates for candidate members of Sco-Cen (top).
The density contours encompass most of the candidates within the fields
marked in Figure~\ref{fig:map} for Upper Sco, Ophiuchus, Lupus, and V1062~Sco.
Candidates that have radial velocities are plotted in diagrams
of $XYZ$ and $UVW$ (middle and bottom) if they are within those four fields
and are kinematic members of those populations. 
In addition, kinematic members of UCL/LCC are plotted
as blue points if they are within the boundary of LCC from \citet{dez99}
and otherwise are plotted as gray points.
}
\label{fig:uvw}
\end{figure}

I have used high-precision photometry and astrometry from Gaia EDR3
to identify candidate members of the stellar populations within the Sco-Cen
complex, which consist of Ophiuchus, Lupus, Upper Sco, V1062~Sco, and UCL/LCC.
The candidates have been used to characterize those populations in terms 
of their IMFs, ages, and space velocities.
The results are summarized as follows:

\begin{enumerate}

\item
For my survey, I have considered the area from $l=283$ to $2\arcdeg$
and $b=-12$ to $35\arcdeg$, which extends beyond the outer boundary of
the Sco-Cen OB subgroups that was defined by \citet{dez99}.
Most of the candidate members of Sco-Cen found in this work are 
located within the latter.

\item
To characterize the kinematics of the populations in Sco-Cen, I began by
identifying candidates for young low-mass stars based on their positions
in Gaia CMDs. The resulting stars exhibit prominent concentrations in 
diagrams of proper motion offset and parallactic distance that correspond to 
the members of Sco-Cen. I applied a Gaussian mixture model to those data to
estimate probabilities of membership in Sco-Cen and the field.
I then used the probable members of Sco-Cen within 
the area in which a given population is concentrated to measure the ranges of 
proper motion offsets and distances exhibited by that population.

\item
I have selected sources from EDR3 that have
$\sigma_{\pi}<1$~mas, proper motion offsets and distances that
overlap with the ranges of values measured for the populations in Sco-Cen,
and photometry that is consistent with membership based on Gaia CMDs.
These criteria produced 10,509 candidate members of Sco-Cen 
(Table~\ref{tab:cand}).
I also identified a second sample of 195 objects from EDR3 that did not 
satisfy the kinematic criteria for membership but that are possible companions
to candidates in the first sample (Table~\ref{tab:comp}).

\item
I have compiled previous measurements of spectral types, Li equivalent widths,
and radial velocities for the candidate members of Sco-Cen, which are available 
for 3169, 1420, and 1740 objects in Table~\ref{tab:cand}, respectively.

\item
Contamination by field stars is reduced by considering only the candidates that
have RUWE$<$1.6, $\sigma_{\pi}<0.5$~mas, $\sigma_{BP}<0.1/\sigma_{RP}<0.1$ or
$\sigma_G<0.1/\sigma_{RP}<0.1$, and locations within the Sco-Cen boundary
from \citet{dez99}, which corresponds to 7272 and 8000 objects for the
two photometric criteria, respectively.
For these samples, I estimate that the contamination from field stars is
$\sim0.6$ and 0.9\%, respectively, and the completeness is $\sim90$\%
for most of the magnitude range of Gaia.
For the range of colors in which eROSITA data are available in
Sco-Cen \citep[$G_{\rm BP}-G_{\rm RP}>1$,][]{sch21}, 
the X-ray detections and limits from eROSITA's first all-sky survey are
consistent with youth for virtually all of the candidates.

\item
I have updated my previous estimates of the intrinsic colors of young
stars and brown dwarfs to include the bands from Gaia EDR3 and
to make use of the candidate members of Sco-Cen from this work 
(Table~\ref{tab:intrinsic}).
For the Sco-Cen candidates that have spectral classifications, I have
estimated extinctions from their color excesses relative to 
the intrinsic colors expected for a given spectral type.
I have estimated both spectral types and extinctions for candidates
that lack spectroscopy by dereddening their observed colors to the sequences
of intrinsic colors in color-color diagrams.

\item
All of the populations in Sco-Cen have histograms of spectral types that
peak near M4--M5, which indicates that they share similar IMFs in terms
of their characteristic masses ($\sim0.15$--0.2~$M_\odot$).
Based on a comparison to deeper spectroscopic surveys of Upper Sco,
the completeness of the Gaia sample of candidates in that region
begins to decrease at spectral types later than M7 ($\lesssim0.06$~$M_\odot$).
Given their ages and distances relative to Upper Sco, the samples in UCL/LCC
and V1062~Sco likely become increasingly incomplete beyond $\sim$M7 and M6,
respectively.
After accounting for incompleteness, I estimate that the Sco-Cen complex
contains nearly 10,000 members with masses above $\sim0.01$~$M_\odot$.

\item
Recent studies have compared the H-R diagrams of low-mass stars
in Sco-Cen and other nearby associations using Gaia DR2,
arriving at ages of 2--6~Myr for groups in Ophiuchus, $\sim$6~Myr 
for Lupus, $\sim$11~Myr for Upper Sco, and $\sim$20~Myr for V1062~Sco
and UCL/LCC \citep{esp20,luh20u,luh20lu}.
Those results are not affected by the data from EDR3
or the new membership classifications in this work.
I have searched for evidence of spatial variations in relative ages among
candidates in the population with the largest spatial extent, UCL/LCC.
The stars in one small corner ($b\lesssim0\arcdeg$)
appear to be $\sim6$~Myr younger than the remainder of UCL/LCC,
which has been noticed in previous analysis of Gaia data \citep{gol18,ker21}.
The portion of UCL/LCC that overlaps with Upper Sco may also have a slight
excess of younger stars. Otherwise, the bulk of UCL/LCC exhibits a uniform
median age with location.

\item
I have calculated $UVW$ space velocities for the Sco-Cen candidates that
have measurements of radial velocities. 
UCL/LCC exhibits evidence of expansion in the form of correlations between
$UVW$ and $XYZ$, respectively. Evidence of this kind was previously reported
for a sample in LCC that was identified with Gaia DR2 \citep{gol18}.
Upper Sco also shows such correlations in $U/X$ and $V/Y$.
The continuous spatial distribution, coherent pattern of expansion, and
relatively uniform median age with location among UCL and LCC candidates
(Figures~\ref{fig:map}, \ref{fig:amap}, \ref{fig:uvw})
indicate that UCL and LCC can be considered to be a single population.

\end{enumerate}

\acknowledgements

I thank Eric Mamajek and Konstantin Getman for comments on the manuscript and 
I thank J\"{u}rgen Schmitt for providing the table of eROSITA sources in Sco-Cen
prior to publication.  This work used data from the European Space Agency 
(ESA) mission Gaia (\url{https://www.cosmos.esa.int/gaia}), processed by
the Gaia Data Processing and Analysis Consortium (DPAC,
\url{https://www.cosmos.esa.int/web/gaia/dpac/consortium}). Funding
for the DPAC has been provided by national institutions, in particular
the institutions participating in the Gaia Multilateral Agreement.
2MASS is a joint project of the University of Massachusetts and IPAC
at Caltech, funded by NASA and the NSF.
WISE is a joint project of the University of California, Los Angeles,
and the JPL/Caltech, funded by NASA. This work used data from the 
NASA/IPAC Infrared Science Archive, operated by JPL under contract
with NASA, and the VizieR catalog access tool and the SIMBAD database, 
both operated at CDS, Strasbourg, France.
SRG is a joint Russian-German science mission supported by the Russian Space
Agency (Roskosmos), in the interests of the Russian Academy of Sciences
represented by its Space Research Institute (IKI), and the Deutsches Zentrum
f\"{u}r Luft- und Raumfahrt (DLR). The SRG spacecraft was built by Lavochkin
Association (NPOL) and its subcontractors, and is operated by NPOL with support
from the Max Planck Institute for Extraterrestrial Physics (MPE). 
The development and construction of the eROSITA X-ray instrument was led by 
MPE, with contributions from the Dr. Karl Remeis Observatory Bamberg \& ECAP
(FAU Erlangen-Nuernberg), the University of Hamburg Observatory, the 
Leibniz Institute for Astrophysics Potsdam (AIP), and the Institute 
for Astronomy and Astrophysics of the University of T\"{u}bingen, with
the support of DLR and the Max Planck Society. The Argelander Institute for 
Astronomy of the University of Bonn and the Ludwig Maximilians Universität 
Munich also participated in the science preparation for eROSITA. 
The Center for Exoplanets and Habitable Worlds is supported by the
Pennsylvania State University, the Eberly College of Science, and the
Pennsylvania Space Grant Consortium.

\clearpage

\LongTables

\begin{deluxetable}{ll}
\tabletypesize{\scriptsize}
\tablewidth{0pt}
\tablecaption{Candidate Members of Sco-Cen from Gaia EDR3\label{tab:cand}}
\tablehead{
\colhead{Column Label} &
\colhead{Description}}
\startdata
GaiaEDR3 & Gaia EDR3 source name \\
GaiaDR2 & Gaia DR2 source name \\
Name & Other source name \\
RAdeg & Right ascension from Gaia EDR3 (ICRS at Epoch 2016.0)\\
DEdeg & Declination from Gaia EDR3 (ICRS at Epoch 2016.0)\\
SpType & Spectral type \\
r\_SpType & Spectral type reference\tablenotemark{a} \\
Adopt & Adopted spectral type \\
l\_EWLi & Limit flag for EWLi \\
EWLi & Average equivalent width of Li \\
r\_EWLi & EWLi reference\tablenotemark{a}\\
pmRA & Proper motion in right ascension from Gaia EDR3\\
e\_pmRA & Error in pmRA \\
pmDec & Proper motion in declination from Gaia EDR3\\
e\_pmDec & Error in pmDec \\
plx & Parallax from Gaia EDR3\\
e\_plx & Error in plx \\
r\_med\_geo & Median of the geometric distance posterior from \citet{bai21}\\
r\_lo\_geo & 16th percentile of the geometric distance posterior from \citet{bai21}\\
r\_hi\_geo & 84th percentile of the geometric distance posterior from \citet{bai21}\\
RVel & Radial velocity \\
e\_RVel & Error in RVel \\
r\_RVel & Radial velocity reference\tablenotemark{b} \\
U & $U$ component of space velocity \\
e\_U & Error in U \\
V & $V$ component of space velocity \\
e\_V & Error in V \\
W & $W$ component of space velocity \\
e\_W & Error in W \\
Gmag & $G$ magnitude from Gaia EDR3\\
e\_Gmag & Error in Gmag \\
GBPmag & $G_{\rm BP}$ magnitude from Gaia EDR3\\
e\_GBPmag & Error in GBPmag \\
GRPmag & $G_{\rm RP}$ magnitude from Gaia EDR3\\
e\_GRPmag & Error in GRPmag \\
RUWE & Renormalized unit weight error from Gaia EDR3 \\
kin & Kinematic population\tablenotemark{c} \\
pos & Position in Sco-Cen\tablenotemark{d} \\
2m & Closest 2MASS source within $3\arcsec$ \\
2msep & Angular separation between Gaia EDR3 (epoch 2000) and 2MASS \\
2mclosest & Is this Gaia source the closest match for the 2MASS source? \\
wise & Closest WISE source within $3\arcsec$ \\
wisesep & Angular separation between Gaia EDR3 (epoch 2010.5) and WISE \\
wiseclosest & Is this Gaia source the closest match for the WISE source? 
\enddata
\tablenotetext{a}{
(1) \citet{hou78};
(2) \citet{can93};
(3) \citet{rod11};
(4) \citet{hou75};
(5) \citet{ria06};
(6) \citet{pec16};
(7) \citet{mam02};
(8) \citet{tor06};
(9) \citet{mur13};
(10) \citet{pec12};
(11) \citet{bow19};
(12) \citet{bus69};
(13) \citet{hil69};
(14) \citet{hou82};
(15) \citet{koe17};
(16) \citet{alc95};
(17) \citet{cov97};
(18) \citet{spe39};
(19) \citet{zuc01};
(20) \citet{fan17};
(21) \citet{man13};
(22) \citet{nes95};
(23) \citet{edw76};
(24) \citet{lev75};
(25) \citet{gah83};
(26) \citet{lev06};
(27) \citet{gag15};
(28) \citet{chen11};
(29) \citet{gar94};
(30) \citet{rei08};
(31) \citet{rai16};
(32) \citet{cor84};
(33) \citet{bra12};
(34) \citet{mur15};
(35) \citet{wic97};
(36) \citet{hou88};
(37) \citet{upg72};
(38) \citet{kra97};
(39) \citet{kah16};
(40) \citet{abt95};
(41) \citet{man17};
(42) \citet{man06};
(43) \citet{luh21sp};
(44) \citet{pre98};
(45) \citet{kun99};
(46) \citet{gla72};
(47) \citet{dez99};
(48) \citet{luh18};
(49) \citet{com13};
(50) \citet{app83};
(51) \citet{hey89};
(52) \citet{hug94};
(53) \citet{her14};
(54) \citet{alc17};
(55) \citet{alc14};
(56) \citet{luh20u};
(57) \citet{riz15};
(58) \citet{jay06};
(59) \citet{all07};
(60) \citet{daw14};
(61) \citet{esp18};
(62) \citet{rui87};
(63) \citet{her77};
(64) \citet{vie03};
(65) \citet{lod06};
(66) \citet{lod08};
(67) \citet{bon14};
(68) \citet{all13b};
(69) \citet{ard00};
(70) \citet{mar04};
(71) \citet{wal94};
(72) \citet{mar10};
(73) \citet{kra09};
(74) \citet{mar94};
(75) \citet{sle08};
(76) \citet{pre02};
(77) \citet{coh86};
(78) \citet{hey90};
(79) \citet{mor01};
(80) \citet{giz02};
(81) \citet{sle06};
(82) \citet{mor11};
(83) \citet{kra15};
(84) \citet{rom12};
(85) \citet{cod17};
(86) \citet{dav19b};
(87) \citet{ans16};
(88) \citet{pre01};
(89) \citet{gue01};
(90) \citet{cru03};
(91) \citet{bej08};
(92) \citet{kra07};
(93) \citet{lod11};
(94) \citet{mue11};
(95) \citet{bra97};
(96) \citet{her09};
(97) \citet{hou99};
(98) \citet{muz14};
(99) \citet{com03};
(100) \citet{muz15};
(101) \citet{blo06};
(102) \citet{man16};
(103) \citet{dav16b};
(104) \citet{sta17};
(105) \citet{lac15};
(106) \citet{sta18};
(107) \citet{coh79};
(108) \citet{pra03};
(109) \citet{eis05};
(110) \citet{mur69};
(111) \citet{mar98a};
(112) \citet{pra02};
(113) \citet{pra07};
(114) \citet{man14};
(115) \citet{mcc10};
(116) \citet{lod15};
(117) \citet{dav16a};
(118) \citet{car06};
(119) \citet{luh05usco};
(120) \citet{mar98b};
(121) \citet{esp20};
(122) \citet{all20};
(123) \citet{cie10};
(124) \citet{clo07};
(125) \citet{luh07};
(126) \citet{mer10};
(127) \citet{eri11};
(128) \citet{wil05};
(129) \citet{mah03};
(130) \citet{gre95};
(131) \citet{luh99};
(132) \citet{bou92};
(133) \citet{alv10};
(134) \citet{str49};
(135) \citet{wil99};
(136) \citet{nat02};
(137) \citet{man15};
(138) \citet{muz12};
(139) \citet{luh97};
(140) \citet{gat06};
(141) \citet{alv12};
(142) \citet{gee07};
(143) \citet{nat06};
(144) \citet{bow14};
(145) \citet{pat93};
(146) \citet{vrb93};
(147) \citet{ryd80};
(148) \citet{esp21};
(149) \citet{son12};
(150) \citet{whi07};
(151) \citet{wic99};
(152) \citet{fin87};
(153) \citet{bia17};
(154) \citet{mag92};
(155) \citet{jam06};
(156) \citet{cut02};
(157) \citet{gut20};
(158) \citet{sch19}.}
\tablenotetext{b}{
(1) \citet{gon06};
Gaia DR2;
(3) \citet{chen11};
(4) \citet{mur13};
(5) \citet{jil06};
(6) \citet{kun17};
(7) \citet{tor06};
(8) \citet{des15};
(9) \citet{sch19};
(10) \citet{mal13};
(11) \citet{gal3};
(12) \citet{son12};
(13) \citet{whi07};
(14) \citet{apo16};
(15) \citet{sou18};
(16) \citet{shk17};
(17) \citet{wic99};
(18) \citet{bud15};
(19) \citet{dah12};
(20) \citet{gal13};
(21) \citet{fra17};
(22) \citet{alc20};
(23) \citet{and83};
(24) \citet{kur06};
(25) \citet{dav16a};
(26) \citet{muz03};
(27) \citet{ric10};
(28) \citet{alo15};
(29) \citet{pra07};
(30) \citet{wan18};
(31) \citet{gut20};
(32) \citet{sul19};
(33) \citet{mat89}.}
\tablenotetext{c}{Parallax and proper motion offset are consistent with
membership in these populations based on the criteria in Figures~\ref{fig:pp1}
and \ref{fig:pp2}:
u = Upper Sco; o = Ophiuchus; l = Lupus; v = V1062 Sco; c = UCL/LCC.}
\tablenotetext{d}{Celestial coordinates within these regions:
o = the Ophiuchus field from \citet{esp18};
u = the triangular field in Upper Sco from \citet{luh20u}, excluding Ophiuchus;
l = the fields encompassing Lupus clouds 1--4 from \citet{luh20lu};
v = a $2\arcdeg$ radius field centered on the V1062~Sco group;
s = the boundary of Sco-Cen from \citet{dez99}.}
\tablecomments{
The table is available in its entirety in machine-readable form.}
\end{deluxetable}

\clearpage

\begin{deluxetable}{ll}
\tabletypesize{\scriptsize}
\tablewidth{0pt}
\tablecaption{Non-kinematic Candidate Companions in Sco-Cen\label{tab:comp}}
\tablehead{
\colhead{Column Label} &
\colhead{Description}}
\startdata
GaiaEDR3 & Gaia EDR3 source name \\
GaiaDR2 & Gaia DR2 source name \\
Name & Other source name \\
RAdeg & Right ascension from Gaia EDR3 (ICRS at Epoch 2016.0) \\
DEdeg & Declination from Gaia EDR3 (ICRS at Epoch 2016.0) \\
SpType & Spectral type \\
r\_SpType & Spectral type reference\tablenotemark{a} \\
Adopt & Adopted spectral type \\
EWLi & Average equivalent width of Li \\
r\_EWLi & EWLi reference\tablenotemark{a}\\
pmRA & Proper motion in right ascension from Gaia EDR3\\
e\_pmRA & Error in pmRA \\
pmDec & Proper motion in declination from Gaia EDR3\\
e\_pmDec & Error in pmDec \\
plx & Parallax from Gaia EDR3\\
e\_plx & Error in plx \\
r\_med\_geo & Median of the geometric distance posterior from \citet{bai21}\\
r\_lo\_geo & 16th percentile of the geometric distance posterior from \citet{bai21}\\
r\_hi\_geo & 84th percentile of the geometric distance posterior from \citet{bai21}\\
RVel & Radial velocity \\
e\_RVel & Error in RVel \\
r\_RVel & Radial velocity reference\tablenotemark{b} \\
Gmag & $G$ magnitude from Gaia EDR3\\
e\_Gmag & Error in Gmag \\
GBPmag & $G_{\rm BP}$ magnitude from Gaia EDR3\\
e\_GBPmag & Error in GBPmag \\
GRPmag & $G_{\rm RP}$ magnitude from Gaia EDR3\\
e\_GRPmag & Error in GRPmag \\
RUWE & Renormalized unit weight error from Gaia EDR3 \\
pos & Position in Sco-Cen\tablenotemark{c} \\
2m & Closest 2MASS source within $3\arcsec$ \\
2msep & Angular separation between Gaia EDR3 (epoch 2000) and 2MASS \\
2mclosest & Is this Gaia source the closest match for the 2MASS source? \\
wise & Closest WISE source within $3\arcsec$ \\
wisesep & Angular separation between Gaia EDR3 (epoch 2010.5) and WISE \\
wiseclosest & Is this Gaia source the closest match for the WISE source? 
\enddata
\tablenotetext{a}{
(1) \citet{tor06};
(2) \citet{ria06};
(3) \citet{hou75};
(4) \citet{can93};
(5) \citet{edw76};
(6) \citet{pec16};
(7) \citet{hou78};
(8) \citet{chen11};
(9) \citet{hil69};
(10) \citet{kra97};
(11) \citet{cor84};
(12) \citet{esp18};
(13) \citet{pre02};
(14) \citet{hou82};
(15) \citet{pre98};
(16) \citet{pec12};
(17) \citet{app83};
(18) \citet{hug94};
(19) \citet{com03};
(20) \citet{luh18};
(21) \citet{bra97};
(22) \citet{hou88};
(23) \citet{riz15};
(24) \citet{mar04};
(25) \citet{luh05usco};
(26) \citet{luh20u};
(27) \citet{cie07};
(28) \citet{mur69};
(29) \citet{wil05};
(30) \citet{eri11};
(31) \citet{pra03};
(32) \citet{son12};
(33) \citet{wic99};
(34) \citet{whi07};
(35) \citet{bia17};
(36) \citet{gut20}.}
\tablenotetext{b}{
(1) Gaia DR2;
(2) \citet{gal3};
(3) \citet{chen11};
(4) \citet{gon06};
(5) \citet{apo16};
(6) \citet{tor06}.}
\tablenotetext{c}{Celestial coordinates within these regions:
o = the Ophiuchus field from \citet{esp18};
u = the triangular field in Upper Sco from \citet{luh20u}, excluding Ophiuchus;
l = the fields encompassing Lupus clouds 1--4 from \citet{luh20lu};
v = a $2\arcdeg$ radius field centered on the V1062~Sco group;
s = the boundary of Sco-Cen from \citet{dez99}.}
\tablecomments{
The table is available in its entirety in machine-readable form.}
\end{deluxetable}

\clearpage

\begin{deluxetable}{lllll}
\tabletypesize{\scriptsize}
\tablewidth{0pt}
\tablecaption{Previous Samples of Candidate Members of Sco-Cen\label{tab:prev}}
\tablehead{
\colhead{Study} &
\colhead{Source of Candidates} &
\colhead{N(Candidates)} &
\colhead{N(EDR3} &
\colhead{N(Candidates}\\
\colhead{} &
\colhead{} &
\colhead{} &
\colhead{parallaxes)\tablenotemark{a}} &
\colhead{in this work)\tablenotemark{b}}}
\startdata
\citet{dez99} & Hipparcos &  521 & 499 & 296 \\
\citet{pre01,pre02} & CMDs, spectra & 166 & 154 & 129 sc/126 u \\
\citet{mam02} & \citet{dez99}, X-rays, & 126 & 118 & 76 \\   
 & \citet{hoo00}, spectra & & & \\
\citet{mar04} & CMDs, spectra & 28 & 26 & 24 sc/21 u \\ 
\citet{chen05} & \citet{dez99} &   40  &    37  &   21 \\
\citet{car06} & various & 218 & 201 & 167 sc/143 u \\
\citet{sle08} & CMDs, spectra & 145 & 139 & 100 \\
\citet{chen11} & \citet{dez99}, spectra &  159  & 155 & 90 \\
\citet{lod11} & \citet{lod11}, spectra & 90 & 87 & 80 sc/73 u \\
\citet{riz11} & Hipparcos+RVs & 436 & 416 & 264 \\
\citet{riz12} & \citet{riz11}+WISE &  146  & 144  & 84 \\
\citet{chen12} & \citet{dez99} & 216 & 203 & 132 \\
\citet{son12} & Hipparcos, $\mu$, X-rays, spectra & 104 & 97 & 62 \\
\citet{luh12} & various & 863 & 804 & 683 sc/590 u \\
\citet{jan12} & \citet{dez99} & 3 & 3 & 2 \\
\citet{jan13} & \citet{dez99} & 138 & 138 & 76 \\
\citet{hin15} & \citet{riz11} & 6 & 6 & 5 \\
\citet{riz15} & CMDs, $\mu$, spectra & 237 & 224 & 207 sc/167 u \\
\citet{pec16} & X-rays, $\mu$, CMDs, spectra & 493 & 465 & 323 \\
\citet{mel17} & various & 162 & 153 & 110 \\
\citet{luh18} & CMDs, $\mu$, spectra & 1631 & 1487 & 1273 sc/1075 u \\
\citet{ros18} & Gaia DR1 & 63 & 61 & 50 sc/39 v \\
\citet{wil18} & Gaia DR1 & 167 & 167 & 127 sc/91 u \\
\citet{gol18} & Gaia DR2 & 1844 & 1843 & 1687 \\
\citet{can19} & Gaia DR2, control sample & 188 & 188 & 169 sc/144 o \\
\citet{can19} & Gaia DR2, common sample & 391 & 391 & 349 sc/260 o \\
\citet{dam19} & Gaia DR2, PMS sample\tablenotemark{c} & 10471 & 10457 & 7913 \\
\citet{dam19} & Gaia DR2, upper MS sample\tablenotemark{c} & 3413 & 3409 & 704 \\
\citet{moo19} & CMDs, best candidates & 118 & 114 & 96 \\
\citet{moo19} & CMDs, good candidates & 348 & 308 & 215 \\
\citet{esp20} & CMDs, $\mu$, Gaia DR2, spectra & 373 & 348 & 308 sc/270 o \\
\citet{luh20u} & CMDs, $\mu$, Gaia DR2, spectra & 1761 & 1580 & 1419 sc/1346 u \\
\citet{tei20} & Gaia DR2 & 128 & 128 & 102 \\
\citet{gal20b} & Gaia DR2 & 137 & 137 & 128 sc/121 lu \\
\citet{luh20lu} & Gaia DR2, on-cloud & 121 & 119 & 111 sc/104 lu \\
\citet{san20} & various & 5 & 4 & 4 sc/3 lu\\
\citet{lov21} & various, Gaia DR2 & 30 & 30 & 21 sc/7 lu \\
\citet{gra21} & Gaia EDR3 & 1281 & 842 & 713 sc/412 o
\enddata
\tablenotetext{a}{Number of candidates from the third column that have 
parallaxes with errors of $<1$~mas from Gaia EDR3.}
\tablenotetext{b}{Number of candidates from the fourth column that are
classified as candidate members of any Sco-Cen population in this work
(Table~\ref{tab:cand}). If two numbers are listed, they refer to the
candidates assigned to any population in Sco-Cen (sc) and the population
that was targeted by the study, either Upper Sco (u), Ophiuchus (o),
Lupus (lu), or V1062~Sco (v).}
\tablenotetext{c}{Excludes candidates that are outside of the survey field
in this work or classified as members of IC~2602 by \citet{dam19}.}
\end{deluxetable}

\clearpage

\begin{deluxetable}{ll}
\tabletypesize{\scriptsize}
\tablewidth{0pt}
\tablecaption{Intrinsic Colors of Young Stars and Brown Dwarfs\label{tab:intrinsic}}
\tablehead{
\colhead{Column Label} &
\colhead{Description}}
\startdata
SpType & Spectral Type \\
BP$-$RP & $G_{\rm BP}-G_{\rm RP}$ for Gaia EDR3 bands \\
G$-$RP & $G-G_{\rm RP}$ for Gaia EDR3 bands \\
RP$-$K & $G_{\rm RP}-K_s$ for Gaia EDR3 and 2MASS bands \\
J$-$H & $J-H$ for 2MASS bands \\
H$-$K & $H-K_s$ for 2MASS bands \\
K$-$3.6 & $K_s-[3.6]$ for 2MASS and Spitzer bands \\
K$-$4.5 & $K_s-[4.5]$ for 2MASS and Spitzer bands\tablenotemark{a} \\
K$-$5.8 & $K_s-[5.8]$ for 2MASS and Spitzer bands \\
K$-$8.0 & $K_s-[8.0]$ for 2MASS and Spitzer bands \\
K$-$24 & $K_s-[24]$ for 2MASS and Spitzer bands\tablenotemark{b} \\
K$-$W1 & $K_s-$W1 for 2MASS and WISE bands \\
K$-$W2 & $K_s-$W2 for 2MASS and WISE bands\tablenotemark{a}\\
K$-$W3 & $K_s-$W3 for 2MASS and WISE bands \\
K$-$W4 & $K_s-$W4 for 2MASS and WISE bands\tablenotemark{b}
\enddata
\tablenotetext{a}{Same values listed for $K_s-[4.5]$ and $K_s-$W2.}
\tablenotetext{b}{Same values listed for $K_s-[24]$ and $K_s-$W4.}
\tablecomments{
The table is available in its entirety in machine-readable form.}
\end{deluxetable}

\clearpage

\begin{deluxetable}{lccccccr}
\tabletypesize{\scriptsize}
\tablewidth{0pt}
\tablecaption{Medians and Standard Deviations of Space Velocities of
Sco-Cen Populations\label{tab:uvw}}
\tablehead{
\colhead{Population\tablenotemark{a}} &
\colhead{$U$} &
\colhead{$V$} &
\colhead{$W$} &
\colhead{$\sigma_U$} &
\colhead{$\sigma_V$} &
\colhead{$\sigma_W$} &
\colhead{$N_*$}\\
\colhead{} &
\colhead{} &
\colhead{(km~s$^{-1}$)} &
\colhead{} &
\colhead{} &
\colhead{(km~s$^{-1}$)} &
\colhead{} &
\colhead{}}
\startdata
Upper Sco & $-$4.6 & $-$16.1 & $-$7.0 & 5.0 & 1.1 & 2.1 & 469 \\
Ophiuchus & $-$5.1 & $-$15.2 & $-$9.4 & 4.2 & 0.8 & 1.6 & 96 \\
Lupus     & $-$2.6 & $-$17.9 & $-$7.1 & 3.4 & 1.5 & 1.2 & 52 \\
V1062~Sco & $-$4.2 & $-$19.4 & $-$4.2 & 2.7 & 0.8 & 0.3 & 27 \\
UCL/LCC   & $-$6.4 & $-$19.9 & $-$5.5 & 4.7 & 3.4 & 1.8 & 542
\enddata
\tablenotetext{a}{These samples are defined by the criteria in 
Section~\ref{sec:clean} and the availability of radial velocity measurements
(Table~\ref{tab:cand}).}
\end{deluxetable}

\clearpage

\end{document}